%% file: main.tex
\documentclass[fleqn,usenatbib]{mnras}

\input{Content/preamble.tex}

\input{Content/header.tex}

\begin{document}
\include{Content/bayesnet}
\label{firstpage}
\pagerange{\pageref{firstpage}--\pageref{lastpage}}
\maketitle

\input{Content/1-Abstract}
\input{Content/2-Introduction}

\input{Content/3-NF}
\input{Content/4-Toy-Model}

\input{Content/5-Cosmological-Datasets}
\input{Content/6-Public-Likelihoods}
\input{Content/7-Code}
\input{Content/8-Conclusion}

\input{Content/acknowledgement}
\input{Content/softwares}
\input{Content/data_availability}

\bibliographystyle{aa}
\bibliography{ads}

\appendix
\input{Content/appendix.tex}

\bsp	
\label{lastpage}
\end{document}

%% file: Content/preamble.tex
\usepackage[T1]{fontenc}
\usepackage{ae,aecompl}

\usepackage{array}
\usepackage{graphicx}	
\usepackage[nice]{nicefrac}
\usepackage{bm}
\usepackage{textcomp}
\usepackage{hyperref}
\usepackage{natbib}
\usepackage{amsmath}
\usepackage{amssymb}
\usepackage{graphicx}
\usepackage[colorinlistoftodos]{todonotes}
\usepackage{xcolor}
\usepackage{multirow}
\usepackage{todonotes}
\usepackage{float}
\usepackage{booktabs}
\usepackage{subfig}
\usepackage{lineno}
\usepackage{seqsplit}
\usepackage{verbatim}
\usepackage{ulem}
\usepackage{calrsfs}
\DeclareMathAlphabet{\pazocal}{OMS}{zplm}{m}{n}

\usepackage{amsfonts}
\usepackage{tabularx}
\usepackage{empheq}
\usepackage[many,most,skins]{tcolorbox}
\usepackage{algorithm}
\usepackage{algpseudocode}

\usepackage{color}
\definecolor{green}{rgb}{0.0,0.5,0.0}
\definecolor{darkblue}{rgb}{0.,0.,0.4}
\definecolor{darkred}{rgb}{0.5,0.,0.}
\hypersetup{urlcolor=darkred, citecolor=blue, linkcolor=blue}

\usepackage{mathptmx}

\usepackage{tikz}

\usepackage{xspace}

\definecolor{ForestGreen}{RGB}{34, 139, 34}

\newcommand{\bb}[1]{\mathbb{#1}}
\newcommand{\bs}[1]{\boldsymbol{#1}}
\newcommand{\tm}[1]{\textrm{#1}}
\newcommand{\sans}[1]{\textsf{#1}}

\newcommand{\mc}[1]{\pazocal{#1}}

\newcommand{\namaster}{\texttt{NaMaster}\xspace}

\newcommand{\planck}{{\sl Planck}\xspace}

\tcbuselibrary{theorems}

\newtcbtheorem[number within=section]{ournote}{~~~}%
{colback=green!3,colframe=green!55!black,fonttitle=\bfseries, separator sign none}{nt}

%% file: Content/header.tex
\title[Normalising Flows in Cosmology]{$\tt{emuflow}$: Normalising Flows for Joint Cosmological Analysis}

\author[A. Mootoovaloo et al]{
Arrykrishna Mootoovaloo $^{1,2}$\thanks{E-mail: arrykrishna.mootoovaloo@physics.ox.ac.uk},
Carlos Garc\'ia-Garc\'ia $^{2}$,
David Alonso $^{2}$,
Jaime Ruiz-Zapatero $^{3,4}$
\\
$^{1}$ LSST-DA Catalyst Fellow\\
$^{2}$ Oxford Astrophysics, Department of Physics, University of Oxford, Denys Wilkinson Building, Keble Road, Oxford, OX1 3RH, UK\\
$^3$ Department of Physics and Astronomy, University College London, Gower Street, London WC1E 6BT, UK. \\
$^4$ Advanced Research Computing Centre, University College London, 90 High Holborn, London WC1V 6LJ, UK.
}

\date{Accepted XXX. Received YYY; in original form ZZZ}

\pubyear{2024}

%% file: Content/bayesnet.tex
%
%
%

\usetikzlibrary{shapes}
\usetikzlibrary{fit}
\usetikzlibrary{chains}
\usetikzlibrary{arrows}

\tikzstyle{latent} = [circle,fill=white,draw=black,inner sep=1pt,
minimum size=20pt, font=\fontsize{10}{10}\selectfont, node distance=1]
\tikzstyle{obs} = [latent,fill=gray!25]
\tikzstyle{const} = [rectangle, inner sep=0pt, node distance=1]
\tikzstyle{factor} = [rectangle, fill=black,minimum size=5pt, inner
sep=0pt, node distance=0.4]
\tikzstyle{det} = [latent, diamond]

\tikzstyle{plate} = [draw, rectangle, rounded corners, fit=#1]
\tikzstyle{wrap} = [inner sep=0pt, fit=#1]
\tikzstyle{gate} = [draw, rectangle, dashed, fit=#1]

\tikzstyle{caption} = [font=\footnotesize, node distance=0] %
\tikzstyle{plate caption} = [caption, node distance=0, inner sep=0pt,
below left=5pt and 0pt of #1.south east] %
\tikzstyle{factor caption} = [caption] %
\tikzstyle{every label} += [caption] %


\newcommand{\factoredge}[4][]{ %
  \foreach \f in {#3} { %
    \foreach \x in {#2} { %
      \path (\x) edge[-,#1] (\f) ; %
    } ;
    \foreach \y in {#4} { %
      \path (\f) edge[->, >={triangle 45}, #1] (\y) ; %
    } ;
  } ;
}

\newcommand{\edge}[3][]{ %
  \foreach \x in {#2} { %
    \foreach \y in {#3} { %
      \path (\x) edge [->, >={triangle 45}, #1] (\y) ;%
    } ;
  } ;
}

\newcommand{\factor}[5][]{ %
  \node[factor, label={[name=#2-caption]#3}, name=#2, #1,
  alias=#2-alias] {} ; %
  \factoredge {#4} {#2-alias} {#5} ; %
}


\newcommand{\gate}[4][]{ %
  \node[gate=#3, name=#2, #1, alias=#2-alias] {}; %
  \foreach \x in {#4} { %
    \draw [-*,thick] (\x) -- (#2-alias); %
  } ;%
}

\newcommand{\vgate}[6]{ %
  \node[wrap=#2] (#1-left) {}; %
  \node[wrap=#4] (#1-right) {}; %
  \node[gate=(#1-left)(#1-right)] (#1) {}; %
  \node[caption, below left=of #1.north ] (#1-left-caption)
  {#3}; %
  \node[caption, below right=of #1.north ] (#1-right-caption)
  {#5}; %
  \draw [-, dashed] (#1.north) -- (#1.south); %
  \foreach \x in {#6} { %
    \draw [-*,thick] (\x) -- (#1); %
  } ;%
}

\newcommand{\hgate}[6]{ %
  \node[wrap=#2] (#1-top) {}; %
  \node[wrap=#4] (#1-bottom) {}; %
  \node[gate=(#1-top)(#1-bottom)] (#1) {}; %
  \node[caption, above right=of #1.west ] (#1-top-caption)
  {#3}; %
  \node[caption, below right=of #1.west ] (#1-bottom-caption)
  {#5}; %
  \draw [-, dashed] (#1.west) -- (#1.east); %
  \foreach \x in {#6} { %
    \draw [-*,thick] (\x) -- (#1); %
  } ;%
}

%% file: Content/1-Abstract.tex
\begin{abstract}
Given the growth in the variety and precision of astronomical datasets of interest for cosmology, the best cosmological constraints are invariably obtained by combining data from different experiments. At the likelihood level, one complication in doing so is the need to marginalise over large-dimensional parameter models describing the data of each experiment. These include both the relatively small number of cosmological parameters of interest and a large number of ``nuisance'' parameters. Sampling over the joint parameter space for multiple experiments can thus become a very computationally expensive operation. This can be significantly simplified if one could sample directly from the marginal cosmological posterior distribution of preceding experiments, depending only on the common set of cosmological parameters. In this paper, we show that this can be achieved by emulating marginal posterior distributions via normalising flows. The resulting trained normalising flow models can be used to efficiently combine cosmological constraints from independent datasets without increasing the dimensionality of the parameter space under study. We show that the method is able to accurately describe the posterior distribution of real cosmological datasets, as well as the joint distribution of different datasets, even when significant tension exists between experiments. The resulting joint constraints can be obtained in a fraction of the time it would take to combine the same datasets at the level of their likelihoods. We construct normalising flow models for a set of public cosmological datasets of general interests and make them available, together with the software used to train them, and to exploit them in cosmological parameter inference.

\end{abstract}

%% file: Content/2-Introduction.tex
\section{Introduction}
\label{sec:introduction}
Data analysis in cosmology is rapidly evolving. With data from past and current experiments such as Planck \citep{2020A&A...641A...6P}, Kilo-Degree Survey (KiDS) \citep{2021A&A...645A.104A}, and Dark Energy Survey (DES) \citep{2022PhRvD.105b3520A}, as well as forthcoming surveys like Euclid \citep{2018LRR....21....2A}, Simons Observatory \citep{2019JCAP...02..056A}, DESI \citep{2024arXiv240403002D}, LSST \citep{2019ApJ...873..111I} amongst others,  there is a growing need for the development of tools that can accelerate the analysis of cosmological data.

\begin{figure*}
    \centering
    \subfloat[$p(\bs{\theta},\bs{\beta}|\bs{x})$]{
        \begin{tikzpicture}[x=1cm,y=2cm]
            \node[obs](c){$\bs{x}$};
            \node[latent, above left=of c](a){$\bs{\theta}$};
            \node[latent, above right=of c](b){$\bs{\beta}$};
            \edge[below=of a]{a}{c};
            \edge[below=of b]{b}{c};

        \end{tikzpicture}
    }
    \hspace{2cm}
    \subfloat[$p(\bs{\theta},\bs{\beta}|\bs{x}_{1},\,\bs{x}_{2})$]{
        \begin{tikzpicture}[x=1cm,y=2cm]
            \node[obs](c){$\bs{x}_{1}$};
            \node[obs, right=1.5cm of c](d){$\bs{x}_{2}$};
            \node[latent, above right=of c](a){$\bs{\theta}$};
            \node[latent, above left=of c](b){$\bs{\beta}$};
            \edge{a}{c};
            \edge{a}{d};
            \edge{b}{c};
        \end{tikzpicture}
    }
    \hspace{2cm}
    \subfloat[$p(\bs{\theta}|\bs{x}_{1},\,\bs{x}_{2})$]{
        \begin{tikzpicture}[x=1cm,y=2cm]
            \node[latent](a){$\bs{\theta}$};
            \node[obs, below left=of a](c){$\bs{x}_{1}$};
            \node[obs, below right=of a](b){$\bs{x}_{2}$};
            \edge[below=of a]{a}{b};
            \edge[below=of a]{a}{c};
        \end{tikzpicture}
    }
    \caption{\label{fig:dags}Directed Acyclic Graphs (DAGs) showing the typical inference problem in cosmology in Panel (a). Panel (b) shows the DAG for a joint analysis in the case where the forward model in experiment 1 also has nuisance parameters, $\bs{\beta}$ and for experiment 2, we have access to an approximate distribution, $p(\bs{\theta}|\bs{x}_{2})$. In Panel (c), we have marginalised over all the nuisance parameters and we have approximate $p(\bs{\theta}|\bs{x}_{1})$ and $p(\bs{\theta}|\bs{x}_{2]})$. Note that we are working with independent datasets, hence there is not link between any two datasets, $\bs{x}_{1}$ and $\bs{x}_{2}$.}
\end{figure*}
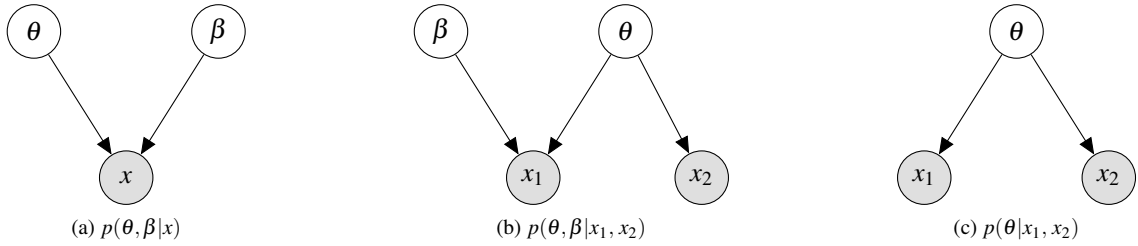

Various techniques have been developed to accelerate computations in cosmology, depending on the tasks being investigated. For instance, generative models have been extensively used in field-level analysis in Cosmology. \citet{2020MNRAS.495.4227K} built a Generative Adversarial Network (GAN) emulator for low-resolution cosmological simulations. \citet{2019MNRAS.487L..24T} explored deep generative models to identify an accurate representation of the large-scale distribution of gas and its temperature. \citet{2024arXiv240807699J} also developed a field-level emulator for large-scale structure structure. Emulators can also be developed at the power spectrum level, though this can be computationally expensive due to the need for numerous forward simulations and power spectrum computations in joint analyses. Recently,  symbolic regression techniques have been used to derive mathematical expressions for linear and non-linear matter power spectra \citep{2023arXiv231115865B, 2024A&A...686A.150B}. In the same spirit, \citet{2022MNRAS.511.1771S, 2021MNRAS.506.4070A} developed power spectrum emulators based on deep learning. On the other hand, \citet{2022A&C....3800508M} developed a Gaussian Process emulator for linear and non-linear matter power spectra, which can also be used for computing weak lensing power spectra.  While these techniques effectively accelerate computations, a major challenge is defining the region of parameter space before building the emulator. A naive broad prior can result in power spectra computations where the cosmological data does not constrain the parameters.

In short, the aim of these Machine Learning methods is to learn an effective model that is able to describe the data, using only the data as input. Data in this context can be any key quantity we are interested in, for example, cosmological samples, bandpowers and other forms of compressed data \citep{2018MNRAS.477.2874A, 2020MNRAS.497.2213M}. On one hand, we have generative models such as GAN, Variational Auto-Encoders (VAE), Normalising Flows, and Gaussian Process which can learn the data directly. On the other hand, we have Simulation-Based Inference (SBI) techniques which learn the parameters of the model by simulating data (see e.g. Bayesian Optimisation for Likelihood-Free Inference -- BOLFI \citep{2018PhRvD..98f3511L}). While the two classes of Machine Learning methods (generative models and SBI) are related by the fact that they both adopt a probabilistic approach, their implementations and goals are different. For example, generative models have a loss function to learn the data distribution, while SBI uses a forward model for the data and the goal is to learn the parameters of the model. In this work, we focus solely on generative model, in particular, normalising flow models. 

In Cosmology, numerous publicly available chains for cosmological and nuisance parameters have been obtained using MCMC-based approaches from different datasets. The question is whether we can exploit these chains (rather than the likelihood and theory prediction codes that generated them) to perform joint analysis of different probes efficiently. A similar concept was investigated by \citet{2017arXiv170403472H, 2017PhRvL.119j1301H}, who utilised publicly available MCMC chains to estimate the marginal likelihood. Moreover, \citet{2023MNRAS.526.4613B} developed a technique which uses normalising flows and kernel density estimators to learn marginal posterior distribution of the scientific parameters in cosmology. In this work, we demonstrate how normalising flows can be employed to learn these marginal probability distributions and subsequently use them to perform joint analyses combining different experiments. This allows us to bypass having to sample computationally expensive and slow joint posterior distributions, as is often the standard approach.

Normalising flows have been used in various applications in Cosmology. For example, \citet{2021MNRAS.505L..95A} combined normalising flow models with nested sampling. Recently, \citet{2024arXiv240412294S} developed a codebase, $\tt{flowZ}$ to estimate the Bayesian Evidence from posterior samples. Normalising flows have also been used in the estimation of Bayesian Evidence via the harmonic mean estimation \citep{2021arXiv211112720M, 2024arXiv240505969P}. Our contributions in this work are as follows. 1) We show that pre-trained normalising flows can be used to effectively and precisely sample joint posterior distributions without increasing the dimensionality of the parameter space or the computational cost of including new complex likelihoods. We show that this is true even for combinations of experiments that are in relatively large tension with each other. 2) We also make many pre-trained models publicly available, together with an API that makes it easy for anyone to include them in their likelihoods.

The paper is organised as follows. In \S\ref{sec:normalising_flows}, we describe the normalising flow procedures before applying the concepts to simple, toy examples in \S\ref{sec:toy_example}. In \S\ref{sec:cosmological_datasets}, we apply the method to infer the cosmological parameters and nuisance parameters. We take two approaches, the first where a normalising flow model is used as a prior and the second case, where we simply use two normalising flows to sample the joint distribution. Furthermore, we use existing publicly available MCMC chains to construct these normalising flow models, as discussed in \S\ref{sec:public_likelihoods}. We also briefly cover how the code works in \S\ref{sec:code}. We discuss our results in \S\ref{sec:results} before concluding in \S\ref{sec:conclusion}.

%% file: Content/3-NF.tex
\section{Normalising Flows}
\label{sec:normalising_flows}

Normalising flows are a class of generative model which transform simple distributions into complex ones via invertible functions. They are efficient tools for density estimation and sampling. In effect, our goal is to employ normalising flows to learn the density function of publicly available MCMC chains of cosmological interest over a few parameters of interest, marginalised over the rest. This is useful in the task of joint analysis as well as specifying a prior before sampling parameters of a model of choice. 

\subsection{Motivation}
\label{sec:motivation}
Suppose we have $N$ experiments, each having its own set of cosmological parameters, $\bs{\theta}_{i}$, nuisance parameters, $\bs{\beta}_{i}$ and data $\bs{x}_{i}$. We are also assuming that the data, $\bs{x}_{i}$ is independent from each other. For simplicity, we will assume that the we have a common set of cosmological parameters across all experiments. Let us also assume that we have samples of $\{\bs{\theta}_{i},\,\bs{\beta}_{i}\}$, which are obtained by sampling the posterior of all parameters in each experiment. The marginalised posterior distribution of the cosmological parameters
\begin{equation}
    p(\bs{\theta}_{i}|\bs{x}_{i})\equiv\int p(\bs{\theta}_{i},\,\bs{\beta}_{i}|\bs{x}_{i})\,\tm{d}\bs{\beta}_{i}.
\end{equation}
 can be obtained by considering only the values of those parameters in the MCMC chain, ignoring the values of $\beta_i$.

Panel (a) of Figure \ref{fig:dags} shows the Directed Acyclic Graph (DAG) for this setup. $\bs{\theta}_{i}$ and $\bs{\beta}_{i}$ are the latent variables and $\bs{x}_{i}$ is the fixed data. We are now interested in finding the joint posterior of the cosmological parameters, $\bs{\theta}$ given the different experiments. In the first case, we can think of a scenario where we want to enforce a more informative prior in the analysis, for example, a case where we have the likelihood for a large-scale structure data $(\bs{x}_{1})$ and we want to use the posterior distribution of cosmological parameters inferred by Planck $(\bs{x}_{2})$ as a prior. In this case,

\begin{equation}
    \label{eq:joint_posterior}
    p(\bs{\theta}|\bs{x}_{1},\,\bs{x}_{2}) = \int p(\bs{x}_{1}|\bs{\theta},\,\bs{\beta})\,p(\bs{\theta}|\bs{x}_{2})\,p(\bs{\beta})\,\tm{d}\bs{\beta}
\end{equation}

\noindent where $p(\bs{x}_{1}|\bs{\theta},\,\bs{\beta})$ is the likelihood of $\bs{x}_{1}$ and $p(\bs{\theta}|\bs{x}_{2})$ being a more informative prior based on the second dataset, $\bs{x}_{2}$. See Panel (b) in Figure \ref{fig:dags} for this setup. In the third scenario, we can also have a case where we simply use nuisance-marginalised models for joint inference of the cosmological parameters. See Panel (c) in Figure \ref{fig:dags}.

If we were to do a joint analysis among the different experiments, the total dimensionality of the problem can become large. For example, if we assume we have $b$ cosmological parameters and each experiment $E_{i}$ has $c_{i}$ nuisance parameters, the total number of parameters is $b+\sum_{i}c_{i}$. Standard sampling schemes such as Metropolis-Hastings may struggle to learn the full posterior distribution of all parameters. Furthermore, as we incorporate more experiments into the analysis, it may become increasingly computationally intensive. Our proposal is to sample the parameters in each experiment (or use publicly available MCMC chains), followed by data fusion, which we discuss in the next section.

\subsection{Data Fusion}
\label{sec:data_fusion}
The process by which multiple data and knowledge is combined together is known as data fusion \citep{2024ITSP...72..275W}. In Machine Learning, a common practice to augment the knowledge of a single model is via federated learning. In this scenario, a model is trained on a set of data, generating a local agent. This agent can further be trained on another similar dataset in a different location, thereby augmenting its knowledge and capability. In a Bayesian setting, each experiment has its own local estimate of its parameters. There are different ways in which this fusion process can be carried out, and its performance of this fusion process depends on the way the priors are used. For example, in non-parametric Bayesian methods such as Gaussian Process (GP), techniques such as Product of Expert (PoE) Bayesian Committee Machine (BCM) have been developed to fuse local estimates of the GP posterior \citep{tresp2000bayesian}. This is particularly helpful in scaling GP to millions of training points. In what follows, we will only cover parametric Bayesian methods, that is, a scenario where we have a forward model with its cosmological and nuisance parameters which are learnt from data. 

In the first case, assuming a common prior, $p(\bs{\theta})$, the joint posterior is given by Bayes' rule as:

\begin{equation}
    p(\bs{\theta}|\bs{x}_{1},\,\bs{x}_{2},\,\ldots\bs{x}_{N})=\dfrac{p(\bs{\theta})\,\prod_{i=1}^{N}p(\bs{x}_{i}|\bs{\theta})}{p(\bs{x}_{1},\,\bs{x}_{2},\,\ldots\bs{x}_{N})}
\end{equation}

\noindent and we have assumed that the joint likelihood can be factorised into their local, individual likelihood as a result of the conditional independence. Under this formalism, where the prior is common across all experiments, each experiment returns a local posterior in the data fusion process. One can also think of a scenario where each experiment has its local prior and we have a global prior for data fusion. In this case, we can write

\begin{equation}
p(\bs{\theta}|\bs{x}_{1},\,\bs{x}_{2},\,\ldots\bs{x}_{N})=Z\,p(\bs{\theta})\,\prod_{i=1}^{N}\dfrac{p(\bs{\theta}|\bs{x}_{i})}{p_{i}(\bs{\theta})}
\end{equation}

\noindent where $Z$ is 

\begin{equation}
  Z=\dfrac{\prod_{i=1}^{N}p(\bs{x}_{i})}{p(\bs{x}_{1},\,\bs{x}_{2},\,\ldots\bs{x}_{N})}
\end{equation}

\noindent and $p_{i}(\bs{\theta})$ is the local prior for each experiment. In the case where the data are independent from each other, $Z=1$. These data fusion techniques are known as \textit{conditionally independent likelihood} (CIL) data fusion \citep{2024ITSP...72..275W}. 

On the other hand, it could also be possible that the $N$ multiple different experiments, have been performed separately and we do not have access to the individual priors but only the local posteriors. One approach to data fusion in this scenario is the Product of Expert (PoE), where these local posteriors are multiplied together to generate a global posterior given by

\begin{equation}
\tilde{p}(\bs{\theta}|\bs{x}_{1},\,\bs{x}_{2},\,\ldots\bs{x}_{N})=c\prod_{i=1}^{N}\,p(\bs{\theta}_{i}|\bs{x}_{i}),
\end{equation}

\noindent where $c$ is some normalisation constant and $\tilde{p}(\bs{\theta}|\bs{x}_{1},\,\bs{x}_{2},\,\ldots\bs{x}_{N})$ is accurate compared to the true posterior, $p(\bs{\theta}|\bs{x}_{1},\,\bs{x}_{2},\,\ldots\bs{x}_{N})$ only when the assumption holds, that is, the case where the priors play an important role generating the global posterior. This data fusion technique is often referred to as the \textit{conditionally independent posterior} (CIP) data fusion \citep{2024ITSP...72..275W}.  Interestingly, in complex data analysis problems which involve expensive and non-linear models, it is highly unlikely that we can find the local, joint posterior of the cosmological parameters only (marginalised over the nuisance parameters), that is, $p(\bs{\theta}_{i}|\bs{x}_{i})$. However, one can try to approximate this local joint posterior of the cosmological parameters via generative modelling frameworks. In this work, given that we have samples of $\bs{\theta}$, the density $p(\bs{\theta}_{i}|\bs{x}_{i})$ is approximated using a normalising flow model. Hence, the approximate joint posterior is:

\begin{equation}
\hat{p}(\bs{\theta}|\bs{x}_{1},\,\bs{x}_{2},\,\ldots\bs{x}_{N})=k\prod_{i=1}^{N}\,p_{\tm{nf}}(\bs{\theta}_{i}|\bs{x}_{i}),
\end{equation}

\noindent where $p_{\tm{nf}}(\bs{\theta}_{i}|\bs{x}_{i})$ is the learned normalising flow model for each experiment and $k$ is just a normalisation constant. Also note that $\hat{p}$ is different from $\tilde{p}$. The assumption that we can combine each individual posterior via the PoE rule, together with approximating the individual posterior with a normalising flow model, can lead to a less accurate joint posterior compared to the true joint posterior, $p(\bs{\theta}|\bs{x}_{1},\,\bs{x}_{2},\,\ldots\bs{x}_{N})$. The approximate joint log-posterior is:

\begin{equation}
\label{eq:nf_log_posterior}
    \tm{log}\,\hat{p}(\bs{\theta}|\bs{x}_{1},\,\bs{x}_{2},\,\ldots\bs{x}_{N})=\sum_{i=1}^{N}\,\tm{log}\,p_{\tm{nf}}(\bs{\theta}_{i}|\bs{x}_{i}) + \tm{log}\,k.
\end{equation}

\noindent If we have access to $\tm{log}\,p_{\tm{nf}}(\bs{\theta}_{i}|\bs{x}_{i})$, we can draw samples from this approximate distribution and we can also compute the log-density. This is crucial because we can then 1) use the log-density for joint analysis and 2) use the learned density as a prior in a completely new cosmological data analysis problem.

Moreover, if we have a pre-trained normalising flow model, it is also possible to combine it in the analysis of a new dataset. If $\bs{x}_{\tm{new}}$ denotes the new dataset and given a pre-trained flow model, $p_{\tm{nf}}(\bs{\theta}|\bs{x}_{\tm{old}})$ of an old dataset, $\bs{x}_{\tm{old}}$, the new posterior due to the joint analysis is: 

\begin{equation}
    p(\bs{\theta},\,\bs{\beta}|\bs{x}_{\tm{new}},\,\bs{x}_{\tm{old}})\propto p(\bs{x}_{\tm{new}}|\bs{\theta},\,\bs{\beta})\,p_{\tm{nf}}(\bs{\theta}|\bs{x}_{\tm{old}})\,p(\bs{\beta})
\end{equation}

\noindent where $\beta$ are the nuisance parameters of the new experiment. $p(\bs{x}_{\tm{new}}|\bs{\theta},\,\bs{\beta})$ and $p(\bs{\beta})$ are likelihood and priors of the nuisance parameters in the new experiment respectively. This is interesting for various reasons. $p_{\tm{nf}}(\bs{\theta}|\bs{x}_{\tm{old}})$ captures all the information about $\bs{x}_{\tm{old}}$ in the cosmological parameters. In this regard, we are enforcing a more informative prior on the cosmological parameters. Moreover, we no longer have to explicitly evaluate the likelihood due to the old data, which can be computationally expensive. 

\subsection{Normalising Flow Theory}
Given $n$ samples of $\bs{\theta}\in\bb{R}^{d}$, a normalising flow provides a simple way to construct a flexible distribution over the cosmological parameters. The idea is to express $\bs{\theta}$ as a transformation, 

\begin{equation}
    \bs{\theta}=f(\bs{z})\hspace{1cm}\bs{z}\sim p(\bs{z})
\end{equation}

\noindent where $\bs{z}\in\bb{R}^{d}$ is sampled from a distribution, $p(\bs{z})$. $p(z)$ is also known as the base distribution. These base distributions are usually simple distributions such as normal distribution or multivariate normal distribution. 

The function $f$, also known as a bijector, has its own set of unknown parameters which we denote as $\phi$. These unknown parameters are learnt via optimisation (see \S\ref{sec:optimisation} below). An important characteristic of $f$ is that it should be invertible and both $f$ and $f^{-1}$ should be differentiable. This also implies that the function is bijective, that is, there is a one-to-one correspondence between elements in the domain of $\theta$ and elements in the codomain of $\bs{z}$. An important property of these types of transformation is that they are also composable. For example, if we have two functions, $f_{1}$ and $f_{2}$, the composition $f_{1}\circ f_{2}$ is also invertible and differentiable. The inverse is given by 

\begin{equation}
    (f_{1}\circ f_{2})^{-1}=f_{2}^{-1}\circ f_{1}^{-1}.
\end{equation}

In general, it is a common practice to combine multiple transformations (bijectors), that is, $f=f_{M}\circ\ldots\circ f_{2}\circ f_{1}$ and each bijector transforms $\bs{z}_{m-1}$ into $\bs{z}_{m}$ and $\bs{z}_{M}=\bs{\theta}$.

\subsubsection{Change of variables}
\label{sec:1d-flow}
Let us consider a 1D example. Suppose we have the continuous random variable, $x$ and its probability density function is $p(\theta)$. In order to change variables, we can write the following:

\begin{equation}
    \int_{\Theta} p(\theta)\,d\theta = \int_{\mc{Z}} p(z)\,\left|\dfrac{dz}{d\theta}\right|\,d\theta
\end{equation}

\noindent where $\Theta$ and $\mc{Z}$ are the support of $\theta$ and $z$ respectively. Therefore, the probability density function of $\theta$ can be written as:

\begin{equation}
    \label{eq:1d_loss}
\tm{log}\,p(\theta)=\tm{log}\,p(z)+\tm{log}\,\left|\dfrac{dz}{d\theta}\right|.
\end{equation}

\noindent In the high dimensional case $(d>1)$, we can write the probability density function as:

\begin{equation}
\label{eq:nd_loss}
\tm{log}\,p(\bs{\theta})=\tm{log}\,p(\bs{z})+\tm{log}\,\left|\tm{det}\left(\dfrac{\partial \bs{z}}{\partial \bs{\theta}}\right)\right|,
\end{equation}

\noindent where $\sans{J}\equiv \frac{\partial \bs{z}}{\partial \bs{\theta}}\in\bb{R}^{d\times d}$ is the Jacobian. Intuitively, we can think of the function, $f$ as warping the space $\bb{R}^{d}$ by moulding the density $p(\bs{z})$ into $p(\bs{\theta})$. The absolute Jacobian determinant term accounts for the volume correction factor. If instead we have a series of transformation, that is, $f=f_{M}\circ\ldots\circ f_{2}\circ f_{1}$, then 

\begin{equation}
\tm{log}\,p(\bs{\theta})=\tm{log}\,p(\bs{z})+\sum_{m=1}^{M}\tm{log}\,\left|\tm{det}\left(\dfrac{\partial \bs{z}_{m-1}}{\partial \bs{z_{m}}}\right)\right|
\end{equation}

\noindent where $\bs{z}_{M}=\bs{\theta}$, $z_{0}=z$ and $z=f_{1}^{-1}\circ\ldots\circ f_{M}^{-1}(\bs{\theta})$. Ideal normalising flows should be expressive, invertible to ensure precise reconstruction of inputs, and have computationally efficient Jacobian determinants to enable quick evaluation and optimization of probability densities, making them amenable to model complex data distributions.

\subsubsection{Optimisation}
\label{sec:optimisation}

Suppose $p_{*}(\theta)$ is the unknown target distribution and we have samples $\{\theta\}_{j=1}^{n}$. The Kullback-Leibler divergence between the target distribution, $p_{*}(\theta)$ and the flow-based model, $p(\bs{\theta}|\phi)$ is:

\begin{equation}
    \mc{L}(\phi) \equiv D_{KL}[p_{*}(\bs{\theta})||p(\bs{\theta}|\phi)]
\end{equation}

\noindent Simplifying the above, we can write the KL-divergence as:
\begin{equation}
\label{eq:kl_divergence}
    \begin{split}
    \mc{L}(\phi)&=-\int p_{*}(\bs{\theta})\tm{log}\,p(\bs{\theta}|\phi)\,d\bs{\theta} + \tm{constant}\\
    &=-\underset{p_{*}(\bs{\theta})}{\bb{E}}\left[\tm{log}\,p(z)+\tm{log}\,\left|\tm{det}\,\dfrac{\partial z}{\partial \bs{\theta}}\right|\right]+ \tm{constant}\\
    &\approx -\dfrac{1}{n}\sum_{j=1}^{n} \left[\tm{log}\,p(z_{j})+\tm{log}\,\left|\tm{det}\,\dfrac{\partial z_{j}}{\partial \bs{\theta}_{j}}\right|\right].
    \end{split}
\end{equation}

\noindent Recall that $z=f^{-1}(\bs{\theta};\phi)$, that is, the bijector $f$, modelled using using neural networks, has unknown parameters $\phi$. Interestingly, minimising the KL-divergence (Equation \ref{eq:kl_divergence}) via the Monte Carlo method is equivalent to fitting the flow model via maximum likelihood estimation. 

In short, once the normalising flow model is trained, this means that we can do two important tasks. First, we can draw samples of $\bs{z}$ from the base distribution and transform them into $\bs{\theta}$ via $\bs{\theta}=f(\bs{z})$. Second, we can calculate the probability density at any point in the $\Theta$ domain via Equation \ref{eq:nd_loss}. The latter is the most crucial aspect for conducting the joint analysis in this work. 

\subsection{Flow Models}
\label{sec:flow_models}
Throughout this work, we will make use of Affine Autoregressive flow model to learn the complex distribution, $p(\bs{\theta})$. Typically, the base distribution, $p(\bs{z})$ is mapped to the $p(\bs{\theta})$ distribution in the forward transformation and is reversed in the backward transformation. In short, in the forward transformation, $\bs{z}\rightarrow \bs{\theta}$ and in the reverse transformation, $\bs{\theta}\rightarrow \bs{z}$. A straightforward example is a scale-location transformation that meets the monotonicity criterion, that is, 

\begin{equation}
    z'=s z + t
\end{equation}

\noindent where $s$ and $t$ are the scale and location parameters respectively. As described in \S\ref{sec:1d-flow}, it is possible to apply a series of $M$ bijective transformations. For a single training point $\bs{\theta}_{j}$, 

\begin{equation}
    \bs{z}_{m}^{(k)}= s_{m}^{(k)}\left(z_{m-1}^{(<k)}\right)\cdot\bs{z}_{m-1}^{(k)}+t_{m}^{(k)}\left(z_{m-1}^{(<k)}\right)
\end{equation}

\noindent where $k$ is the index denoting the $k^{th}$ dimension of the vector $\bs{\theta}$ and $m$ is the $m^{th}$ transformation. Moreover, $z_{0}= z$ and $\bs{z}_{M}=\bs{\theta}$. Both $s$ and $t$ are parameterised by neural networks. Moreover, for each transformation $m$ and each training point, $j$, the absolute Jacobian determinant is:

\begin{equation}
    \tm{log}\,\left|\tm{det}\left(\sans{J}_{m}\right)\right|=\sum_{k=1}^{d}\tm{log}\,\left|s_{m}^{(k)}\left(z_{m-1}^{(<k)}\right)\right|
\end{equation}

\noindent and the total absolute determinant due to $M$ transformations is simply the sum of the above, that is, 

\begin{equation}
\label{eq:tot_log_det}
\tm{log}\,\left|\tm{det}\left(\sans{J}_{\tm{tot}}\right)\right|=\sum_{m=1}^{M}\tm{log}\,\left|\tm{det}\left(\sans{J}_{m}\right)\right|.
\end{equation}

This type of affine transformation yields a lower triangular Jacobian matrix, allowing the determinant to be computed as the product of its diagonal elements. This process is repeated for all training points, and the loss is calculated using Equation \ref{eq:kl_divergence}. By minimising the KL-divergence, the neural network parameters, $\phi$ can be optimised. In this work, we employ three transformations, each with a dense architecture consisting of three layers, each having 32 hidden units and using the $\tt{tanh}$ activation function.

If we are given a test point, $\bs{\theta}_{\tm{test}}$ and we want to compute the log-density, it is mapped to $\bs{z}$ via the reverse transformations, $f^{-1}_{m}$ and the log-density of the base distribution is calculated. Moreover, using Equation \ref{eq:tot_log_det}, the log-determinant of the Jacobian is computed, followed by the computation of $\tm{log}\,p(\bs{\theta}_{\tm{test}})$. On the other hand, if we want to draw $N$ samples from the normalising flow model, $N$ samples from the base distributions are drawn and the forward transformations are applied to map them to the $\bs{\theta}$ space.

\subsection{Metrics}
\label{sec:metrics}
In order to quantify the difference between the approximate posterior (built upon the normalising flow models) and the known joint posterior, we can compute metrics related to the statistics of marginalised posterior distribution of each parameter in 1D. For example, we can compare the mean, $\mu_{\tm{nf}}$ obtained using the normalising flow models with the expected mean, $\mu$, that is, 

\begin{equation}
    \delta_{\mu} = \left|\dfrac{\mu - \mu_{\tm{nf}}}{\mu}\right|. 
\end{equation}

\noindent This essentially gives a measure of how accurate the samples from the normalising flow model is compared to known samples from the joint posterior. Moreover, we also compare the width of the distribution, that is, the standard deviation using

\begin{equation}
    \delta_{\sigma} = \dfrac{|\sigma - \sigma_{\tm{nf}}|}{\sigma}. 
\end{equation}

\noindent This effectively quantifies whether the precisions of the two set of samples are comparable. We can also take the difference of the means divided by the quadrature sum of the errors in both experiments, that is,

\begin{equation}
    \delta_{q} = \dfrac{|\mu - \mu_{\tm{nf}}|}{\sqrt{\sigma^{2} + \sigma_{\tm{nf}}^{2}}}. 
\end{equation}

Note that these metrics apply only in the 1D case. As discussed by \citet{2021MNRAS.505.6179L}, there is no universal method for quantifying the differences in multi-dimensional parameter spaces. However, the lower the values of these metrics, the better the reconstructed posteriors with the normalising flow models. The question is whether we can have a metric to assess the similarity of the two distributions (known joint posterior and the joint posterior due to the normalising flow models). One could use the analytic expression for the KL divergence or the  Bhattacharyya distance or other variant such as the Jensen-Shannon (JS) divergence to quantify the similarity of the two distributions. However, these expressions apply only in the multivariate normal case and in our case, we can have non-Gaussian-like posteriors - see Figure \ref{fig:des_y3_kids_1000} in the appendix. 

In our case, we have samples from the posteriors and one option is to use the Maximum Mean Discrepancy (MMD) which is commonly used in GANs \citep{2018arXiv180101401B}. The idea is to embed the distributions into the reproducing kernel Hilbert space (RKHS) and measure the distance between the means in the embedded space, that is, 

\begin{equation}
\label{eq:mmd}
    \tm{MMD}(p,\hat{p}) = \left\Vert \mu_{p}-\mu_{\hat{p}}\right\Vert_{\mc{H}}.
\end{equation}

For example, we can use the Gaussian kernel to embed the samples and compute the distance using Equation \ref{eq:mmd}. However, the resulting distance can be inconsistent due to its dependence on the bandwidth of the Gaussian kernel. Instead, one can use the energy distance which does not depend on any hyperparameter at all. It is given by:

\begin{equation}
  D^{2}(p,\hat{p})= 2\bb{E}\left\Vert \bs{\theta}-\hat{\bs{\theta}}\right\Vert - \bb{E}\left\Vert \bs{\theta}-\bs{\theta}'\right\Vert - \bb{E}\left\Vert \hat{\bs{\theta}}-\hat{\bs{\theta}}'\right\Vert
\end{equation}

\noindent where $\left\Vert \cdot \right\Vert$ is the Euclidean norm. $\bs{\theta}$ and $\hat{\bs{\theta}}$ are samples from $p$ and $\hat{p}$ respectively. The energy metric is inspired by Newton's concept of gravitational potential energy, where the potential energy becomes zero when the gravitational centres of two particles coincide. If the two distributions ($p$ and $\hat{p}$) are exactly the same, then $D=0$. However, as argued by \citet{rizzo2016energy}, the above distance statistics is not standardized and in order to interpret the value, one can use 

\begin{equation}
    \tilde{D}(p,\hat{p})= \dfrac{2\bb{E}\left\Vert \bs{\theta}-\hat{\bs{\theta}}\right\Vert - \bb{E}\left\Vert \bs{\theta}-\bs{\theta}'\right\Vert - \bb{E}\left\Vert \hat{\bs{\theta}}-\hat{\bs{\theta}}'\right\Vert}{2\bb{E}\left\Vert \bs{\theta}-\hat{\bs{\theta}}\right\Vert},
\end{equation}

\noindent where $0\leq \tilde{D}(p,\hat{p}) \leq 1$. A low value of $\tilde{D}(p,\hat{p})$ indicates a higher degree of similarity between the distributions. There is no clear consensus on what constitutes a good energy metric; relative comparisons are generally more meaningful. We take a random set of 3000 samples from each distribution and compute $\tilde{D}(p,\hat{p})$ in the two experiments we have performed. The results are quoted in Table \ref{tab:metrics}.

%% file: Content/4-Toy-Model.tex
\section{Toy Examples}
\label{sec:toy_example}
In this section, we will look at two examples (1D and 2D) to demonstrate 1) how the normalising flow model works and 2) how it can be used to sample the joint posterior of parameters of interest without requiring the original datasets and likelihoods. 

\subsection{1D Distribution}
\label{sec:1d_distribution}

Let us consider a mixture of three Gaussian distributions, 

\begin{equation}
\label{eq:mixture_gaussians}
    p(\theta)=\sum_{i=1}^{3}w_{i}\mc{N}(\mu_{i},\,\sigma^{2}_{i})
\end{equation}

\noindent where $\sum_{i=1}^{3}w_{i}=1$. The means and the standard deviations assumed are $\bs{\mu}=(-1.0,\,0.5,\,0.0)$ and $\bs{\sigma}=(0.25,\,0.50,\,0.10)$. We will assume a uniform distribution as the base distribution, that is, $p(z)=\mc{U}[0,\,1]$. We will define the bijector, $z=f^{-1}(\theta)$ as a linear combination of cumulative density function, $\Phi(\mu,\,\sigma^{2})$, of the normal distributions, that is,

\begin{equation}
    z=\sum_{c=1}^{C}w_{c}\Phi(\theta;\,\mu_{c},\,\sigma^{2}_{c})
\end{equation}

\noindent where $C$ is the number of components which we are free to choose. In this case, we fix $C=3$. In total, there are 9 parameters in this model:  $\{\mu_{c},\,\sigma_{c},\,w_{c}\}_{c=1}^{3}$. The derivative of the above function with respect to $\theta$ is analytical and can be written as:

\begin{equation}
    \dfrac{dz}{d\theta}=\sum_{c=1}^{C}w_{c}\,\mc{N}(\theta;\,\mu_{c},\,\sigma^{2}_{c})
\end{equation}

\begin{figure}
\noindent \begin{centering}
\includegraphics[width=0.45\textwidth]{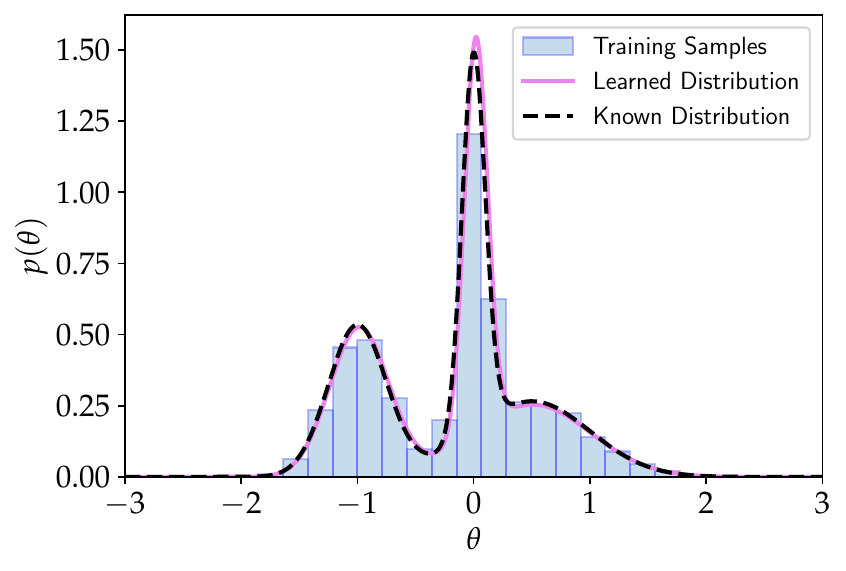}
\par\end{centering}
\caption{\label{fig:1d_mixture_gaussians}The plot shows the training samples in blue. They are generated using a mixture of three normal distributions, with means, $[-1.0, 0.5, 0.0]$ and standard deviations $[0.25, 0.50, 0.10]$. Therefore, $p(\theta)=\sum_{i=1}^{3}w_{i}\mc{N}(\mu_{i}, \sigma_{i})$, where $w_{i}=\frac{1}{3}$ is fixed. The probability distribution learned by the normalising flow model is shown in violet, while the dashed black curve shows the known distribution. The flow model accurately captures the distribution of the generated samples. See explanation in \S\ref{sec:1d-flow} for further details on the implementation.}
\end{figure}

We draw 10000 samples from the true underlying distribution (Equation \ref{eq:mixture_gaussians}) and fit for the 9 parameters $\{\mu_{c},\,\sigma_{c},\,w_{c}\}_{c=1}^{3}$ by maximising Equation \ref{eq:1d_loss} (equivalent to minimising the negative of the log density) using all the samples. The final probability density learnt is shown in violet in Figure \ref{fig:1d_mixture_gaussians}. The same technique explained for this toy 1D example can be extended to higher dimensional scenarios, with the exception, that the bijector is now composed of neural network blocks to accommodate for more expressive functions. 

\subsection{Gaussian Linear Model and Banana Posterior}
\label{sec:glm_banana}

\begin{figure}
\noindent \begin{centering}
\includegraphics[width=0.45\textwidth]{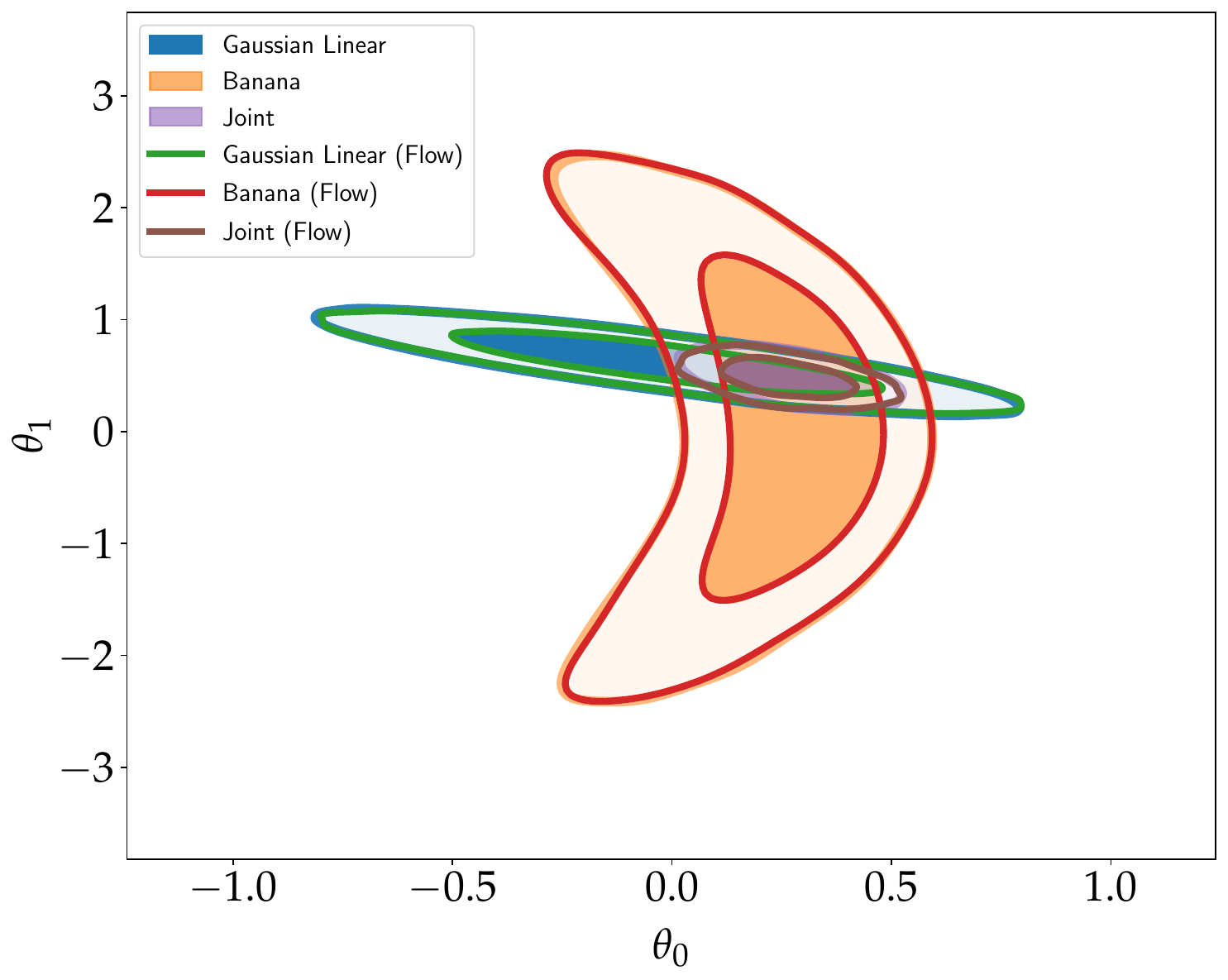}
\par\end{centering}
\caption{\label{fig:banana_gl_joint}The figure shows the joint posterior distribution of a Gaussian posterior, obtained from a Gaussian Linear Model and a banana posterior. See \S\ref{sec:glm_banana} for implementation details. The blue and orange colors show the posteriors of the two parameters $\theta_{0}$ and $\theta_{1}$, sampled using MCMC, for the Gaussian Linear Model and the banana respectively. The normalising flows are built using these samples and are shown in green and red respectively. The purple shaded region shows the joint distribution using the individual likelihoods, while the brown contour shows the joint distribution using only the normalising flow models.}
\end{figure}

Before building the normalising flow models for complex posteriors in Cosmology, in this example, we show that we can recover the joint posterior distribution, solely using the normalising flow models, without needing the original datasets and likelihood functions.

Let us consider a Gaussian Linear Model (GLM) of the form $f(x;\bs{\theta})=\theta_{0}+\theta_{1}x$. The fiducial point is $\bs{\theta} = [0.25,\,0.25]$ and we generate data points, $y=f(x;\bs{\theta})+\epsilon$, where $\epsilon\sim \mc{N}(0, 1)$. The next function we consider is a banana-shape posterior, whose functional form is $g(\bs{\theta})=\theta_{0}+0.1\theta_{1}^{2}$. As in the GLM, we generate 100 data points, that is, $w=g(\bs{\theta})+\epsilon$. We assume independent normal priors -- with mean centred on zero and standard deviation equal to one -- on the parameters, $\bs{\theta}$. $p(\bs{y}|\bs{\theta})$ and $p(\bs{w}|\bs{\theta})$ are the Gaussian likelihoods for the two datasets, $\bs{y}$ and $\bs{w}$ respectively.

We sample the posterior of the individual data using $\tt{EMCEE}$ \citep{2013PASP..125..306F} and the joint posterior between $\theta_{0}$ and $\theta_{1}$ is shown in Figure \ref{fig:banana_gl_joint}. The blue one corresponds to the GLM while the orange one corresponds to the banana function. As explained in \S\ref{sec:motivation}, we can also do a joint analysis by combining the likelihoods, that is, 

\begin{equation}
    p(\bs{\theta}|\bs{y},\,\bs{w})\propto p(\bs{y}|\bs{\theta})\,p(\bs{w}|\bs{\theta})\,p(\bs{\theta})
\end{equation}

\noindent This joint posterior is shown in Figure \ref{fig:banana_gl_joint} in purple. Given that we have MCMC samples, that is, $\bs{\theta}_{y}\leftarrow p(\bs{\theta}|\bs{y})$ and $\bs{\theta}_{w}\leftarrow p(\bs{\theta}|\bs{w})$, we can use a subset of these samples to fit a normalising flow model to learn the posterior probability distribution independently. We first apply an affine transformation (rotation and translation) to the original samples, that is, 

\begin{equation}
    \bs{\theta}' = \sans{L}^{-1}(\bs{\theta} - \bs{\mu}),
\end{equation}

\noindent where $\sans{L}$ is the Cholesky factor of the covariance of the samples, $\bs{\theta}$ and $\bs{\mu}$ is the mean of the samples. We then choose independent normal distributions, $p(\bs{z};\,\bs{z}_{\tm{med}},\,\sigma')$, where $\bs{z}_{\tm{med}}$ and $\sigma'$ are the median and standard deviation of the $\bs{\theta}'$ samples respectively. We choose the median because posteriors may have complex shapes, for example, the banana posterior, and hence the median is a more representative measure of central tendency. Note that the normalising flow will output the density of $p(\bs{\theta'})$. We can draw samples from the original distribution by first sampling, $\bs{\theta}'$, followed by applying the inverse transformation, that is, $\bs{\theta}=\sans{L}\bs{\theta}'+\bs{\mu}$. We can calculate the density using

\begin{equation}
    p(\bs{\theta}) = \dfrac{p(\bs{\theta}')}{\left|\tm{det}\,(\sans{L})\right|}
\end{equation}

\noindent where $\left|\tm{det}\,(\sans{L})\right|$ is the absolute determinant of the Cholesky factor and accounts for the volume correction factor. 

Once the normalising flow models are trained, we can draw samples in green and red for the GLM and banana functions respectively in Figure \ref{fig:banana_gl_joint}. By inspecting the blue and green posteriors for the GLM, and the orange and red posteriors for the banana function, it is evident that the flow models excel at learning the distribution. Our next task is to learn the joint distribution using the normalising flow models only, that is, $p(\bs{\theta}|\bs{y},\,\bs{w})\propto p_{\tm{nf}}(\bs{\theta}|\bs{y})\,p_{\tm{nf}}(\bs{\theta}|\bs{w})$. We sample this joint distribution using $\tt{EMCEE}$ and the joint posterior is shown in brown in Figure \ref{fig:banana_gl_joint}. This is an important result as it demonstrates the ability to recover the joint distribution using the flow models alone, with MCMC samples from individual experiments serving as training points.

%% file: Content/5-Cosmological-Datasets.tex
\section{Validation}\label{sec:validation}
\label{sec:cosmological_datasets}
  An obvious application of emulated marginalised posterior distributions using normalising flows is the possibility of using them as effective priors in the joint analysis of past experiments with new datasets without having to sample the full parameter space of the past datasets (including their nuisance parameters). The aim of this section is validating this approach, applying it to two real datasets that are in mild tension with one another (and for which, therefore, the normalising flow emulator must be able to capture the outskirts of the distributions). We present the two datasets used (a large suite of large-scale structure data from \cite{2021JCAP...10..030G}, and CMB data from \planck), the methodology used for parameter inference, and the results of this validation exercise.

  \begin{figure}
    \noindent
    \begin{centering}
    \includegraphics[width=0.45\textwidth]{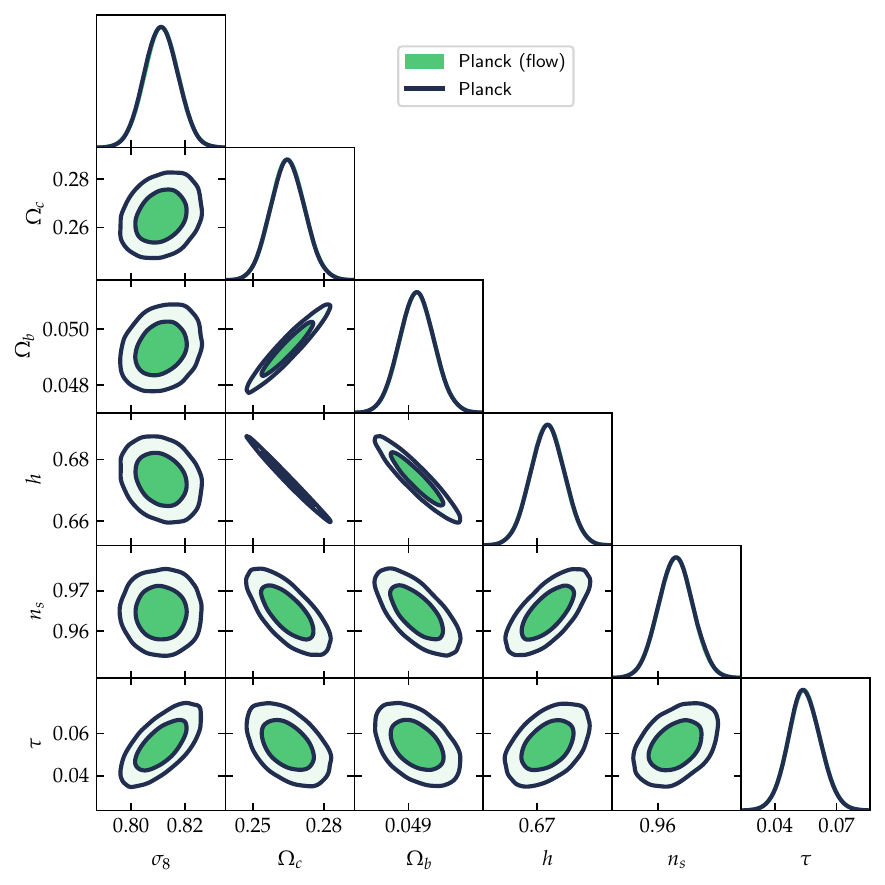}
    \par\end{centering}
    \caption{\label{fig:planck}Figure showing the joint posterior of the cosmological parameters only(marginalised over the nuisance parameters) from the P18 dataset. The green contours correspond to the samples obtained using the normalising flow model and the black contours are the original samples.}
  \end{figure}

  \begin{figure*}
    \centering
    \subfloat[CGG21 and P18 flow as prior]{{\includegraphics[width = 0.45\textwidth]{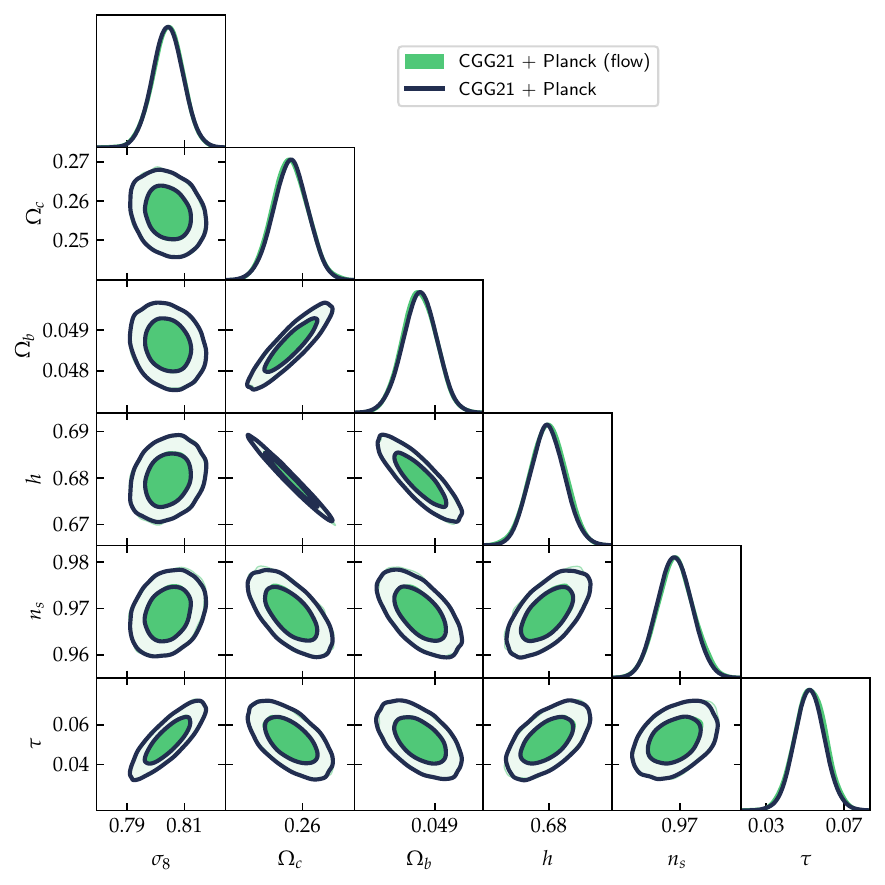}}} 
    \qquad
    \subfloat[CGG21 flow and P18 flow]{{\includegraphics[width = 0.45\textwidth]{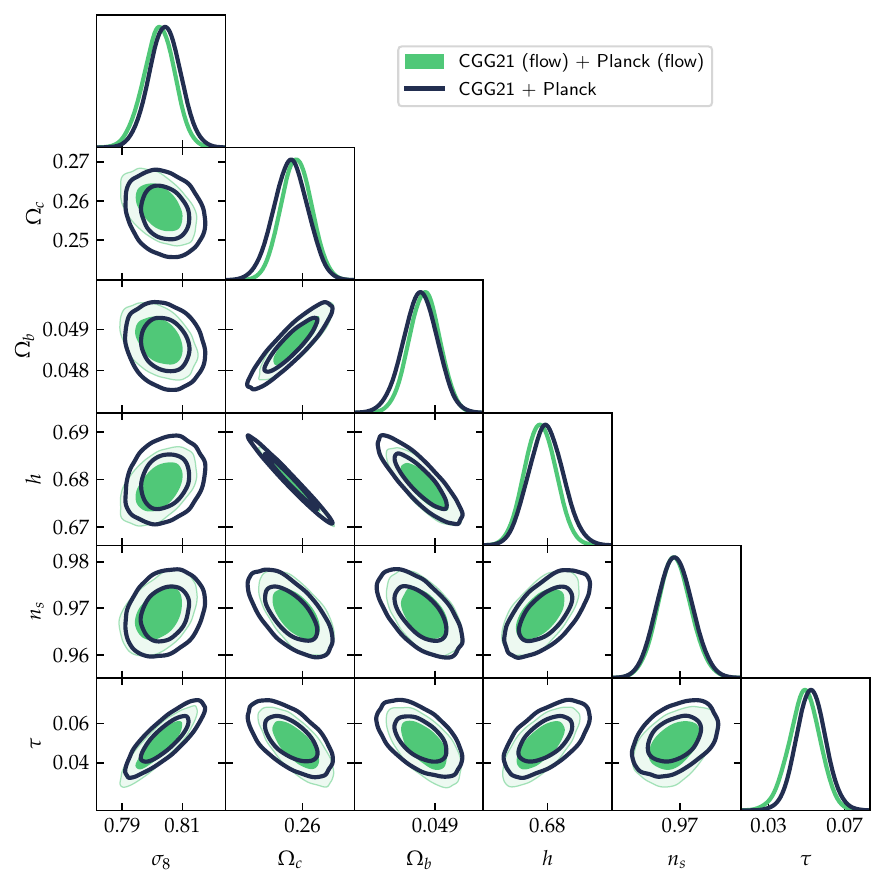}}}
    \caption{\label{fig:triangle_plot_joints}Panel (a) shows the joint posterior of the cosmological parameters where the P18 normalising flow is used as a prior in the analysis. Panel (b) shows the posterior in the case where the posterior is sampled using the local posterior due to the CGG21 and P18 datasets, where each density is learnt by the normalising flow model. In both plots, the green contours correspond to the posterior due to the normalising flow models, whereas the black contours are the known posterior as obtained by \citet{2021JCAP...10..030G}.}
  \end{figure*}

\subsection{Data}\label{sec:cgg21_planck}
  We make use of two cosmological datasets. The first one is the 2018 \planck dataset (P18 hereafter), combining auto and cross-correlations between temperature and $E$-mode polarisation. We use data from the {\tt Commander} component separation algorithm for the low-$\ell$ data ($2\leq\ell\leq29$) and the {\tt plick} likelihood for the high-$\ell$ data in the range $30\leq\ell\leq2508$ for $TT$ and $30\leq\ell\leq1996$ for $TE$ and $EE$  \citep{Planck:2019nip}. We also include the CMB lensing auto-correlation, considering the range of scales $8 \leq L \leq 400$ \citep{Planck:2018lbu}. We generate MCMC chains using the public likelihoods implemented in {\tt Cobaya} \citep{2005.05290}, marginalising over nuisance parameters with priors as recommended in \cite{Planck:2019nip}. 

  The second dataset consists of a large combination of projected large-scale structure data, analysed in \citet{2021JCAP...10..030G} (CGG21 from now on). Details about the dataset itself and the model used to analyse it are described in detail in \citet{2021JCAP...10..030G}, and we provide only a short summary here. The dataset is composed of the angular power spectrum of the auto- and cross-correlation of galaxy clustering from DES-Y1 RedMaGiC \citep{1507.05460}, DESI Legacy Survey (DELS) (specifically, the sample defined in \cite{2010.00466}) and eBOSS QSO targets \citep{2007.08998}; weak lensing from the DES-Y1 \textsc{METACALIBRATION} sample \citep{1708.01533} and KiDS-1000 \citep{2021A&A...645A.104A}; and CMB lensing from \planck 2018 \citep{1807.06210}. We only consider correlations between pairs of datasets that have non-zero sky overlap, we discard cross-correlations between different galaxy clustering samples, as well as the CMB lensing auto-correlation. To avoid modelling the covariance between DELS and DES RedMaGiC, all the region declination $\delta < -36\,\deg$ in DELS was removed.
  
  The CGG21 dataset was analysed starting at the catalogue level in order to ensure a consistent measurement and analysis pipeline for the different datasets, and to properly account for the correlations between them. Maps of all large-scale structure tracers were generated using with a {\tt HEALPix} resolution parameter $N_{\rm side} = 4096$, and power spectra and their covariance were estimated using the pseudo-$C_\ell$ approach with \namaster \citep{1809.09603} method and the Gaussian part of the covariance with the improved Narrow Kernel Approximation. Power spectra involving galaxy clustering were analysed on scales $k<0.15\,{\rm Mpc}^{-1}$, and weak lensing data was analysed with a small-scale cut $\ell<2048$. Different large-scale cuts were used for different tracers depending on the evidence of large-scale systematics in them.

  The model used to analyse these data in \cite{2021JCAP...10..030G} includes 40 different nuisance parameters: linear biases for all clustering samples (11 parameters), multiplicative biases for all comic shear samples (9 parameters), redshift distribution shifts for all photometric samples (18 parameters), intrinsic alignment amplitude and evolution parameters (2 parameters). Of these, the multiplicative bias parameters and the redshift distribution shifts are marginalised over analytically using the methods described in \cite{2007.14989,2301.11978}, leaving 13 nuisance parameters to marginalise over at the level of the likelihood.

  As in \cite{2021JCAP...10..030G}, we compute the linear matter power spectrum $P_{\rm mm}$ with the Boltzmann code \texttt{CLASS} \citep{1104.2933} and the non-linear correction with \texttt{halofit}\citep{1208.2701}. The kernels and angular power spectra are computed with the Core Cosmology Library (\texttt{CCL}) \citep{1812.05995}.

  Finally, we use CMB data from Planck 2018. We use the public likelihoods, as implemented in \texttt{Cobaya}. In particular, we use the auto- and cross-correlations of the temperature ($T$) and polarisation ($E$) fields. We use the data from the \texttt{Commander} component separation algorithm for the low-$\ell$ data ($2 \leq \ell \leq 29$) and the \texttt{plik} likelihood for the high-$\ell$ data in the range $30 \leq \ell \leq 2508$ for $TT$ and $30 \leq \ell \leq 1996$ for $TE$ and $EE$ \cite{Planck:2019nip}. We marginalise over the nuisance parameters with priors as recommended in \citep{Planck:2019nip}. Finally, we include the CMB lensing auto-correlation power spectrum, considering the range of scales $8 \leq L \leq 400$ \citep{2020A&A...641A...6P}.

  We will denote the CGG21 dataset as $\bs{x}_{\tm{cgg21}}$ and the Planck data as $\bs{x}_{\tm{p18}}$.

\subsection{Parameter Inference}\label{sec:fd_planck_parameter_inference}
  We run MCMC chains to sample the individual likelihoods of the CGG21 and P18 datasets. 
  For the CGG21 analysis, we have a set of 13 nuisance parameters, which we denote as $\bs{\beta}$. In addition to these, we consider five $\Lambda$CDM cosmological parameters, which we denote $\bs{\theta}$:
  \begin{equation*}
    \bs{\theta}=\{\sigma_{8},\,\Omega_{c},\,\Omega_{b},\,h,\,n_{s}\}.
  \end{equation*}
  For the P18 analysis, there are $\sim 20$ nuisance parameters, and the cosmological parameters also include the optical depth to reionisation $\tau$. We use top-hat, uninformative flat priors on all cosmological parameters.
  
  Having generated chains for each individual experiment, we train two normalising flows to recover the marginalised posterior distribution of cosmological parameters in each dataset. With both flows at hand, we can now consider three different setups to sample the joint likelihood of both experiments:
  \begin{itemize}
    \item The exact joint cosmological constraints, obtained from the product of the P18 and CGG21 likelihoods marginalised over nuisance parameters:
    \begin{align}\nonumber
      p(\bs{\theta},\tau|\bs{x}_{\tm{p18}},\bs{x}_{\tm{cgg21}})\propto \int &d\bs{\beta}_{\tm{p18}}\,d\bs{\beta}_{\tm{cgg21}}\times\\\nonumber
      &p(\bs{x}_{\tm{p18}}|\bs{\theta},\tau,\bs{\beta}_{\tm{p18}})p(\bs{x}_{\tm{cgg21}}|\bs{\theta},\bs{\beta}_{\tm{cgg21}})\times\\
      &p(\bs{\beta}_{\tm{p18}})p(\bs{\beta}_{\tm{cgg21}})p(\bs{\theta},\tau).
    \end{align}
    \item The product of the true CGG21 likelihood with the normalising flow for the marginalised P18 distribution, using it as an effective prior:
    \begin{equation}
    \begin{split}
      p(\bs{\theta},\tau|\bs{x}_{\tm{p18}},\bs{x}_{\tm{cgg21}})&\propto p_{\tm{nf}}(\bs{\theta},\,\tau|\bs{x}_{\tm{p18}}) \times \\ & \int d\bs{\beta}_{\tm{cgg21}}\,p(\bs{x}_{\tm{cgg21}}|\bs{\theta},\bs{\beta}_{\tm{cgg21}})p(\bs{\beta}_{\tm{cgg21}}).
      \end{split}
    \end{equation}
    This setup has the advantage that we do not need to evaluate the relatively expensive theoretical model and likelihood of the P18 data, and that we do not increase the dimensionality of the parameter space by combining it with CGG21.
    \item The product of normalising flows for both CGG21 and P18:
    \begin{equation}
      p(\bs{\theta},\,\tau|\bs{x}_{\tm{cgg21}},\,\bs{x}_{\tm{p18}})\propto \dfrac{p_{\tm{nf}}(\bs{\theta}|\bs{x}_{\tm{cgg21}})\,p_{\tm{nf}}(\bs{\theta},\,\tau|\bs{x}_{\tm{p18}})}{p(\bs{\theta},\,\tau)}.
    \end{equation}
  \end{itemize}

  As described in \cite{2021JCAP...10..030G}, analysed within the model described above, the CGG21 dataset exhibits tension with P18 in the value of $S_8\equiv\sigma_8\,\sqrt{\Omega_m/0.5}$ at the level of 3.5$\sigma$. The normalising flow models must therefore capture the tails of both distributions with sufficient accuracy for them to recover the correct joint posterior distribution. Quantifying this is the goal of the next section.

  \subsection{Validation results}\label{sec:results}

    Let us start by studying the performance of the normalising flows generated for each experiment individually. The 1D and 2D marginalised constraints obtained from the P18 chain (black contours) and from its normalising flow (green contours) are shown in Figure \ref{fig:planck}. The flow is able to recover the marginalised posterior with excellent accuracy. A similarly accurate flow density is recovered for the CGG21 data. We used $2\times 10^{4}$ training points, randomly selected from the MCMC chain, to train the normalising flow model. Since the models depend only on five or six parameters training them is very quick, taking $\sim 2$ minutes on a desktop computer. After training, generating samples from the trained flow is nearly instantaneous.

    Having access to the flow-based emulators for the individual marginalised posterior distributions, we can now ask ourselves whether these are accurate enough to be used in the joint analysis of different experiments. In a typical scenario, we want to combine the cosmological constraints obtained from a previous legacy experiment with data from a new experiment, for which the model depends on a given set of nuisance parameters, in addition to the common cosmological parameters. In this case, the emulated marginalised posterior for the legacy experiment may be used as an effective prior in the posterior distribution of the new data, avoiding the need to extend the full model parameter space to include the nuisance parameters of the legacy experiment. The combination of CGG21 and P18 allows us test this approach in a particularly challenging scenario. As discussed in \cite{2021JCAP...10..030G}, the CGG21 data displays a $\sim3.5\sigma$ tension in the value of $S_8$ with respect to P18. Thus, a combination of both datasets making use of the P18 emulated posterior is only possible if the flow is able to describe the tails of the distribution accurately. If the method is able to succeed in this case, its application to joint analyses of experiments that are in better agreement with one another would only be more reliable. The result is summarised in the left panel of Figure \ref{fig:triangle_plot_joints}. The figure shows, in black, the exact marginal posterior distribution on the $\Lambda$CDM cosmological parameters, found by sampling the product of the CGG21 and P18 likelihoods, including all their nuisance parameters. In turn, the constraints obtained by sampling the CGG21 likelihood using the P18 trained flow as an effective prior are shown in green. The latter approach is able to recover the exact posterior at very high accuracy. This is further quantified in Table \ref{tab:metrics}, which lists the distance metrics $(\delta_\mu,\delta_\sigma,\delta_q)$. Posterior means are recovered at sub-percent level, and widths at the level of a 2-3\%. The energy distance metric $\tilde{D}$ is very close to zero, signifying an excellent agreement between both distributions.

\begin{table*}
\caption{\label{tab:metrics}The different metrics (as discussed in \S\ref{sec:metrics}) for the different analyses performed. The columns correspond to $\delta_{\mu}$, $\delta_{\sigma}$ and $\delta_{q}$ for the joint analysis involving the CGG21 and P18 datasets. Columns 2 to 4 present the metrics when the P18 flow is used as a prior in the analysis, while the final three columns show the metrics when two normalising flows are used. The last row gives the energy metric, which measures the degree of similarity of two distributions from their respective samples.}
\renewcommand{\arraystretch}{1.5}
\noindent \begin{centering}
\begin{tabular}{|l|>{\centering}p{1.5cm}|>{\centering}p{1.5cm}|>{\centering}p{1.5cm}||>{\centering}p{1.5cm}|>{\centering}p{1.5cm}|>{\centering}p{1.5cm}|}
\cline{2-7} \cline{3-7} \cline{4-7} \cline{5-7} \cline{6-7} \cline{7-7} 
\multicolumn{1}{l|}{} & \multicolumn{3}{c||}{P18 flow as a prior} & \multicolumn{3}{c|}{CGG21 flow and P18 flow}\tabularnewline
\cline{2-7} \cline{3-7} \cline{4-7} \cline{5-7} \cline{6-7} \cline{7-7} 
\multicolumn{1}{l|}{} & $\delta_{\mu}$ & $\delta_{\sigma}$ & $\delta_{q}$ & $\delta_{\mu}$ & $\delta_{\sigma}$ & $\delta_{q}$\tabularnewline
\hline 
Amplitude of density fluctuations, $\sigma_{8}$ & 0.000 & 0.006 & 0.002 & 0.002 & 0.021 & 0.273\tabularnewline
\hline 
CDM density, $\Omega_{\textrm{cdm}}$ & 0.000 & 0.015 & 0.017 & 0.006 & 0.107 & 0.254\tabularnewline
\hline 
Baryon density, $\Omega_{\textrm{b}}$ & 0.000 & 0.007 & 0.021 & 0.002 & 0.107 & 0.187\tabularnewline
\hline 
Hubble parameter, $h$ & 0.000 & 0.027 & 0.011 & 0.002 & 0.096 & 0.252\tabularnewline
\hline 
Scalar spectral index, $n_{s}$ & 0.000 & 0.031 & 0.041 & 0.000 & 0.036 & 0.012\tabularnewline
\hline 
Optical Depth to Reionisation, $\tau$ & 0.005 & 0.025 & 0.023 & 0.067 & 0.052 & 0.318\tabularnewline
\hline 
Energy distance, $\tilde{D}(p,\,\hat{p})$ & \multicolumn{3}{c||}{$\sim0.001$} & \multicolumn{3}{c|}{$\sim0.01$}\tabularnewline
\hline 
\end{tabular}

\par\end{centering}
\end{table*}

    The right panel of Figure \ref{fig:triangle_plot_joints} shows the result of using the product of both trained flow models in order to obtain joint cosmological constraints. In this case we see that, although the product of emulated posteriors yields contours that are very similar to the true posterior (shown again in black), the differences between them are clearly visible. As quantified in Table \ref{tab:metrics}, these correspond to shifts in the posterior means of up to $\sim0.3\sigma$, and the energy distance metric $\tilde{D}$ grows by a factor $\sim10$ with respect with the previous case (while still staying relatively low). As discussed above, this is not entirely surprising, since the level of tension between these datasets requires both normalising flow models to describe the distribution outskirts accurately.

    The ability to use trained flow models to describe the marginal posterior of legacy datasets leads to significant gains in computational speed for joint analyses. The exact joint posterior found by sampling the product of the P18 and CGG21 likelihoods -- which includes the union of their full parameter spaces -- using {\tt Cobaya} takes $\sim 24$ days using two HPC nodes until convergence is reached at the level of $R-1\leq0.02$. If instead we use the P18 flow model as a prior to the CGG21 likelihood, the joint posterior is recovered after only $\sim 6$ days, corresponding to a factor $\sim4$ speed-up. On the other hand, if we simply use the two flow models, joint constraints can be obtained in under $\sim 15$ minutes by generating 120{,}000 MCMC samples using $\tt{EMCEE}$ on a desktop computer.

    In addition to this study, we also perform additional tests, exploring other experiment combinations. These are described in Appendix \ref{sec:additional_tests}. In particular we investigate the joint analysis of KiDS-1000 and DES-Y3 (\S\ref{sec:desy3_kids_1000}), and the combination of P18 and DES-Y1 (\S\ref{sec:des_y1_planck}). These cases correspond to posterior distributions exhibiting lower levels of tension than the CGG21+P18 case described here, but displaying markedly less ``Gaussian'' marginalised constraints. The qualitative results obtained above are confirm and remain valid in these cases.

Although we have only considered pairwise combinations of experiments, normalising flow models may be used to facilitate the combination of arbitrary numbers of independent datasets. Care should be taken, however, that no individual-dataset posterior incorporates informative priors on the cosmological parameters, and that any non-informative but non-flat priors are corrected for. For instance, using Bayes' theorem:
\begin{equation}
  p(\bs{\theta},\,\bs{\beta}|\bs{x}_{1},\bs{x}_{2},\ldots,\bs{x}_{N})\propto p(\bs{x}_{1}|\bs{\theta},\bs{\beta})p(\bs{\theta})p(\bs{\beta})\left[\prod_{i=2}^{N}\dfrac{p_{\tm{nf}}(\bs{\theta}|\bs{x}_{i})}{p(\bs{\theta})}\right],
\end{equation}
where $\bs{x}_{1}$ and $\bs{\beta}$ are the data and nuisance parameters of the main experiment, respectively, $\bs{\theta}$ are the shared cosmological parameters, and $p(\theta)$ is the uninformative prior on $\theta$ assumed in all cases.

%% file: Content/6-Public-Likelihoods.tex
\section{Public Likelihoods}
\label{sec:public_likelihoods}
In addition to the above analyses, we have also created a ``Cosmological Zoo of Normalising Flow Models", where we have taken public MCMC samples, processed them in such as way that we retain only the cosmological parameters, effectively marginalising over the nuisance parameters. We then train and store the respective normalising flow, which can then be used for two purposes as investigated in \S\ref{sec:cosmological_datasets}, (1) to perform joint analysis using the flow models only and (2) to use them as a prior in a cosmological analysis (see Equation \ref{eq:joint_posterior}) using independent datasets. We will briefly describe the emulated chains below. The software used to generate these emulated chains is made publicly available, and users can easily expand on this set of public emulated likelihoods.

\subsection{Planck 2018}
\label{sec:planck_public}
A variety of MCMC chains were made publicly available by the \planck collaboration. For concreteness and simplicity, we use the $\tt{base\_plikHM\_TTTEEE\_lowl\_lowE}$ MCMC samples \citep{2020A&A...641A...6P}. 

As outlined in Section 2 of \citet{2020A&A...641A...6P}, the samples were produced using a base $\Lambda$CDM cosmological model evaluated with $\tt{CAMB}$ \citep{2000ApJ...538..473L}. This particular experiment focused on the TT, TE, and EE power spectra for $\ell > 30$, along with the low-$\ell$ likelihoods. To estimate the TT, TE, and EE power spectra, the $\tt{Plik}$ high multipole likelihood was applied, employing a Gaussian approximation. For modelling the small-scale non-linear matter power spectrum, $\tt{HMcode}$ \citep{2015MNRAS.454.1958M, 2016MNRAS.459.1468M} was used. Six cosmological parameters and numerous nuisance parameters were sampled, with derived parameters such as $\sigma_{8}$ and $H_{0}$ also being recorded. 

\subsection{DES Y3}
We build a normalising flow for the marginalised cosmological posterior of the DES Y3 ``$3\times 2$-point analysis'' using the full 5000 $\tm{deg}^{2}$ of imaging data \citep{2022PhRvD.105b3520A}. The data vector includes the two-point functions for galaxy clustering, galaxy-galaxy lensing and cosmic shear:
$$
\bs{d}\equiv \{\hat{w}^{i}(\theta),\,\hat{\gamma}^{ij}_{t}(\theta),\,\hat{\zeta}^{ij}_{\pm}(\theta)\}
$$
where only autocorrelations are considered for galaxy clustering. After applying the necessary scale cuts, we are left with 462 elements in the data vector.  \citet{2022PhRvD.105b3520A} employed two cosmological models, namely, $\Lambda$CDM and $w$CDM, which have a total of 31 and 32 parameters, respectively, including 25 nuisance parameters. We emulate the MCMC chains with fixed neutrino mass. 

\subsection{KiDS-1000}
For KiDS-1000, we generate normalising flow emulators for the $\Lambda$CDM chains obtained from different analysis choices.

\citet{2021A&A...645A.104A} made use of different two-point statistics for modelling the cosmic shear data. In particular, correlation functions, band power spectra, and COSEBIs were used, marginalising over 7 nuisance and astrophysical parameters. The posterior is sampled using nested sampling \citep{2019OJAp....2E..10F}. The forward cosmological model entails the calculation of the linear matter power spectrum with $\tt{CAMB}$ \citep{2000ApJ...538..473L} and the non-linear contribution using $\tt{HMcode}$ \citep{2015MNRAS.454.1958M, 2016MNRAS.459.1468M}. The different statistics yield different constraints on the cosmological parameters (see Figure 6 from \citet{2021A&A...645A.104A}). We provide normalising flow models, for each of these three types of two-point statistics.

On the other hand, \citet{2021A&A...646A.140H} performed a joint cosmological analysis using KiDS-1000, BOSS \citep{2015ApJS..219...12A} and 2dFLenS \citep{2016MNRAS.462.4240B} data. The flat $\Lambda$CDM forward model has 20 parameters, 5 cosmological parameters and 15 nuisance and astrophysical parameters, and the posterior was sampled using nested sampling. The posterior distribution displays a $\sim 3\sigma$ with \planck in the value of $S_8$. The pre-trained normalising flow based on the MCMC samples of this experiment is also made available.

Finally, \citet{2023OJAp....6E..36D} performed a hybrid analysis using DES Y3 and KiDS-1000 cosmic shear data only. The goal was to investigate and compare different modelling strategies employed by the DES and KiDS teams separately. When the DES data only is used, the model has 17 parameters (6 cosmological and 11 nuisance and astrophysical parameters). On the other hand, when modelling the KiDS-1000 data, there are 14 parameters (6 cosmological and 8 nuisance and astrophysical parameters). We also train normalising flow models for each individual experiment in this setup.

\subsection{ACT DR4}
We also used the public MCMC chains for ACT (DR4 TT+TE+EE) to build a normalising flow models for the joint cosmological parameters. \citet{2020JCAP...12..047A} performed different analyses which include joint analyses with WMAP and Planck separately. In this work, we use the MCMC samples generated using ACT data alone including $TT$, $TE$ and $EE$ binned CMB bandpowers. The $\Lambda$CDM model, with 6 cosmological parameters, is used in the parameter estimation task and two derived parameters ($H_{0}$ and $\sigma_{8}$) are also recorded.

\subsection{SDSS}
Recently, \citet{2021PhRvD.103h3533A} conducted extensive cosmological analyses using data from the completed Sloan Digital Sky Survey (SDSS), encompassing SDSS, SDSS II, the Baryon Oscillation Spectroscopic Survey (BOSS), and the extended BOSS (eBOSS). These datasets allow for the extraction of various cosmological measurements, including baryon acoustic oscillations (BAO). \citet{2021PhRvD.103h3533A} explored several joint analyses involving cosmic shear, CMB temperature and polarization, supernovae, BAO, and other data sources to gain deeper insights of different cosmological models. We use the CMB+BAO and CMB+BAO+SN chains to train two separate normalising flow models. Training each flow model with 20,000 MCMC samples took only about 2 minutes. The original analyses involved 6 cosmological parameters and 20 nuisance parameters, which were marginalised over using $\tt{Cobaya}$.

%% file: Content/7-Code.tex
\section{Software}
\label{sec:code}
The software used to generated normalising flow emulators of marginalised cosmological posteriors is publicly available. We describe this software briefly here.

In the first step, the user processes the data in such a way that only the samples for a reduced set of cosmological parameters are retained. Specifically, we use the 5 $\Lambda$CDM parameters
\begin{equation}
    \bs{\theta}=\{\sigma_{8},\,\Omega_{c},\,\Omega_{b},\,h,\,n_{s}\}.
\end{equation}
A configuration file is then created for this specific experiment, containing the experiment name, the learning rate, the number of optimisation steps and the number of training points to be used for training the normalising flow model. If we do not specify the number of training points, all the samples will be used for training. However, we recommend using around $\sim 2\times 10^{4}$ if available. The model can be trained both for a single configuration or a combination of configurations (different learning rates, number of optimisation steps, number of training points). After the training procedure, the code stores the trained model and plot the loss curve and the projected 1D and 2D distributions of the original samples and the samples generated by the normalising flow model.

Once the models are trained, users may use them in joint cosmological analyses. We provide example code that describes how to use the trained normalising flows as part of an MCMC run using $\tt{EMCEE}$ \citep{2013PASP..125..306F} (although nothing in the software limits its use to this particular sampler).

%% file: Content/8-Conclusion.tex
\section{Conclusion}
\label{sec:conclusion}
In this work, we have explored how normalising flow models can be used to learn the cosmological posterior distribution marginalised over nuisance parameters. Once trained and stored, these models can be used for various purposes, such as generating large numbers of samples of cosmological parameters, or calculating the log-density at any point in parameter space, which can lead to a significant simplification in the way data from independent experiments are combined.

We have performed and assessed different experiments of how the pre-trained models can be used. Using the PoE approach explained in \S\ref{sec:data_fusion}, any $N$ normalising flow models can be combined by multiplying them together, effectively summing the log-posterior and hence enabling fast sampling of the joint posterior of $N$ experiments. We have investigated this approach using two methods: (1) applying the flow model as a prior in a likelihood analysis, and (2) using two or more normalising flow models to sample the joint posterior. These techniques have been thoroughly tested with a combination of large scale structure datasets and Planck, as well as with other experiments like DES Y3 and KiDS-1000 in Appendix \ref{sec:additional_tests}.

Even in the case where we have a significant degree of tension between two sets of parameters in the joint analysis of P18 and CGG21, we have shown that we can recover the marginalised posterior distribution of the cosmological parameters, with good precision $(\delta_{\sigma}\lesssim 0.11)$ and accuracy $(\delta_{\mu} \lesssim 0.07)$. When the flow model is used as a prior, the joint posterior is closer (lower $\delta_{\mu}$, $\delta_{\sigma}$, $\delta_{q}$ and $\tilde{D}$) to the posterior of the full analysis. This is expected since we are coupling only one approximate density (compared to two or more) with the likelihood. 

To test the method further, we have also used other datasets to sample the joint posterior. For example, a joint analysis using the DES Y3 and KiDS-1000 data using their flow models results in a comparable posterior with the known joint posterior distribution of the cosmological parameters. If we extend this further and add the contribution due to the P18 normalising flow model, the parameters shift in the expected directions (see Appendix \ref{sec:additional_tests}).

In addition to using our own MCMC samples for the above experiments, we have also processed, trained and stored normalising flow models for public MCMC chains. A few highlighted here are \planck, DES Y3, KiDS-1000, ACT and SDSS. The software used to train and exploit these normalising flows is written is a simple way and it should be straightforward for users to implement and train new models. Importantly, training new models, or extending the cosmological model under study (e.g. to include neutrino masses or dynamical dark energy), is both straightforward and computationally inexpensive, as long as sufficient training data exists in the form of MCMC samples.



%% file: Content/acknowledgement.tex
\section*{Acknowledgement}
We thank Dr. Zafiirah Hosenie for reviewing this manuscript and providing useful feedback. AM is supported through the LSST-DA Catalyst Fellowship project; this publication was thus made possible through the support of Grant 62192 from the John Templeton Foundation to LSST-DA. DA and CGG acknowledge support from the Beecroft Trust. JRZ is supported by UK Space Agency grants ST/W001721/1. We made extensive use of computational resources at the University of Oxford Department of Physics, funded by the John Fell Oxford University Press Research Fund.

%% file: Content/softwares.tex
\section*{Softwares}
\label{sec:softwares}
The following $\tt{Python}$ libraries have been used as part of this project: $\tt{flowtorch}$ \citep{WebbFlowtorch}, $\tt{GetDist}$ \citep{2019arXiv191013970L}, $\tt{SciPy}$ \citep{2020NatMe..17..261V}, $\tt{NumPy}$ \citep{harris2020array}, $\tt{pandas}$ \citep{jeff_reback_2020_3630805}, $\tt{Cobaya}$ \citep{2013PhRvD..87j3529L}, $\tt{JAX-COSMO}$ \citep{2023OJAp....6E..15C}, $\tt{PyTorch}$ \citep{2019arXiv191201703P} and $\tt{hydra}$ \citep{Yadan2019Hydra}. 

%% file: Content/data_availability.tex
\section*{Data Availability}
The code and part of the data products underlying this article can be found at: \href{https://github.com/Harry45/emuflow/}{https://github.com/Harry45/emuflow/}. The processed data and pre-trained normalizing flows are also made public in the folders $\tt{samples/}$ and $\tt{flows/}$ respectively.

%% file: Content/appendix.tex
\section{Additional Tests}
\label{sec:additional_tests}
To further test the method described in this work, we perform additional tests using other datasets. First, in \S\ref{sec:desy3_kids_1000}, we consider the KiDS-1000 and DES Y3 cosmic shear analysis, where the posteriors are broad, with strong constraints on only relatively few cosmological parameters. In \S\ref{sec:des_y1_planck}, we use the normalising flow model for the \planck public MCMC samples as a prior in a joint cosmological analysis and compare the results with the case where two normalising flow models are used jointly. Using these datasets, we reach conclusions consistent with those from the main cosmological examples (CGG21 and Planck 2018) discussed in this work. This consistency highlights the robustness and reliability of the method presented in this paper.

\subsection{KiDS-1000 and DES Y3}
\label{sec:desy3_kids_1000}
To test the performance of the pre-trained normalising flows, we also look into inferring cosmological parameters only from large scale structure datasets, which currently yields broad posteriors and a non-Gaussian joint posterior in the $\sigma_{8}$ and $\Omega_{\tm{m}}$ plane. \citet{2024arXiv240313794G} carried out a joint analysis of the KiDS-1000, DES Y3 and HSC-DR1 cosmic shear datasets. To model the non-linear matter power spectrum, \citet{2024arXiv240313794G} use the $\tt{Baccoemu}$ emulator whilst taking advantage of the baryonification procedure implemented in the algorithm. The forward model for either dataset consists of different nuisance parameters (shifts in the redshift distributions and multiplicative biases). 

\begin{figure}
\noindent \begin{centering}
\includegraphics[width=0.45\textwidth]{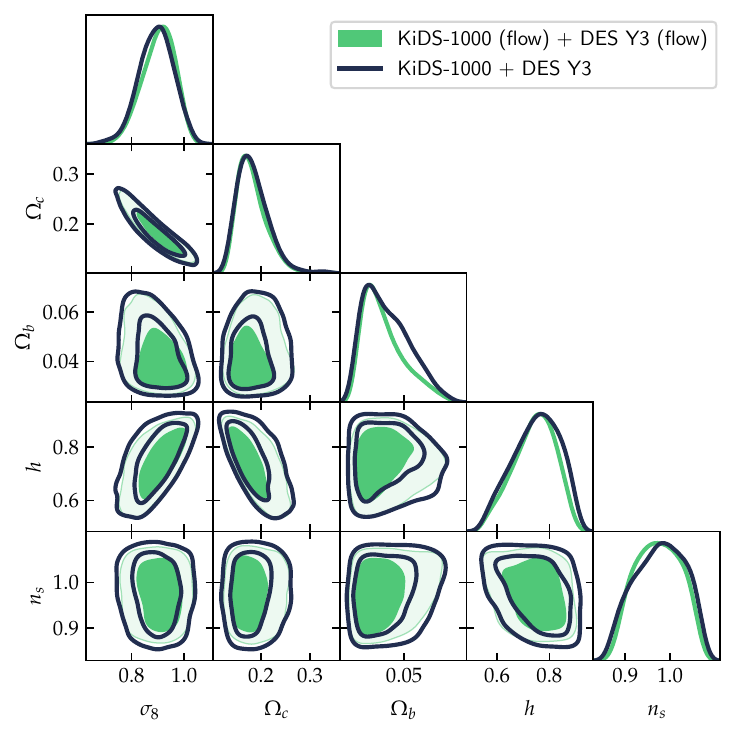}
\par\end{centering}
\caption{\label{fig:des_y3_kids_1000}The marginalised posterior distribution of the cosmological parameters using KiDS-1000 and DES Y3 data. The contours in black shows the result due to the exact evaluation of the likelihoods while the contours in green correspond to the case where we jointly sample the individual normalising flow models for DES Y3 and KiDS-1000.}
\end{figure}

We build two normalising flows based on the cosmological samples, $\bs{\theta}$ and we compute the joint posterior using $\tt{EMCEE}$. This posterior is compared to the MCMC samples obtained from the full run, where the nuisance parameters in the forward modelling both datasets are marginalised over. Results are shown in Figure \ref{fig:des_y3_kids_1000}. The two flows when sampled together, are able to capture the non-Gaussian banana-shape posterior in the $\sigma_{8}$ and $\Omega_{c}$ plane and there are only very mild differences in the posterior distributions. This difference might also arise due to the overlapping area of the DES Y3 and KiDS-1000 surveys, which is not fully accounted for in the original joint analysis.

\subsection{DES Y1 and Planck}
\label{sec:des_y1_planck}

Similar to the analysis done in \S\ref{sec:cosmological_datasets}, in this section we would like to investigate if we can use a pre-trained model as part of a cosmological analysis, that is, we want to simulate a scenario in which we aim to explore the constraints obtained with a new experiment in combination with a previous independent dataset (for which we have built a normalising flow model). As an example, we will use bandpowers data for DES Y1 galaxy clustering and cosmic shear dataset, as well as the \planck public constraints. The forward model is described in \citet{2024arXiv240604725M}. It has five cosmological parameters, similar to the ones used for building the normalising flow model discussed in the main text. The forward model also has 20 nuisance parameters, which we would like to marginalise over. We will also use the pre-trained normalising flow for the Planck 2018 ($\tt{base\_plikHM\_TTTEEE\_lowl\_lowE}$). As discussed in \S\ref{sec:motivation}, we only have to marginalise over the nuisance parameters, $\bs{\beta}$, for DES Y1.
\noindent where we have introduced the approximate posterior built using the normalising flow model for Planck 2018. Note that the normalising flow model should be weighted by the prior in the analysis.

To sample the joint posterior, we have used the Metropolis-hastings sampler implemented in $\tt{Cobaya}$ \citep{2013PhRvD..87j3529L}. We have set the specifications as follows: the number of samples is $5\times 10^{5}$ and the Gelman-Rubin convergence criterion is $R-1=0.01$. The sampler will stop once either of these criteria is met. A total of $215 \times 10^{3}$ MCMC samples were generated and the Gelman-Rubin convergence criterion was met. Sampling takes around 5 hours in this setup, roughly similar to what it would take if we were to sample the cosmological and nuisance parameters in DES Y1.

On the other hand, we also use each individual flow (DES Y1 flow and Planck flow) to find the joint posterior distribution of the cosmological parameters. Sampling the joint in this case is quick and takes $\sim 15$ minutes on a desktop computer.

\begin{figure}
\noindent \begin{centering}
\includegraphics[width=0.45\textwidth]{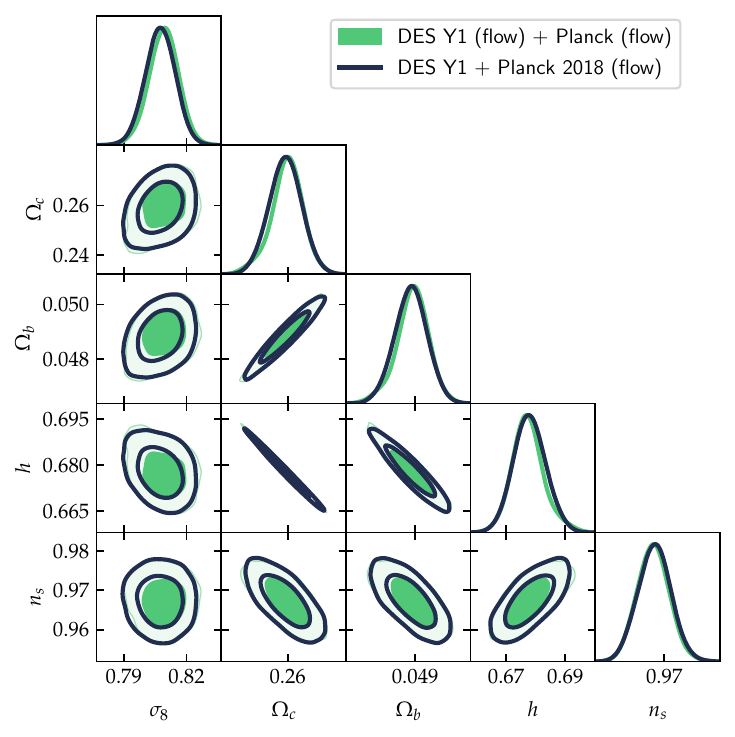}
\par\end{centering}
\caption{\label{fig:des_y1_planck}The marginalised posterior distribution of the cosmological parameters in the joint analysis of the DES Y1 and Planck 2018. For the latter, we use the public MCMC chains for Planck 2018 to train a normalising flow model, which is then used as a prior in conjunction with the DES Y1 likelihood. This result is shown in black. In a separate analysis, we also simply use two normalising flows (DES Y1 and Planck 2018) to sample the joint posterior, which is shown in green. There are mild differences between the two distributions (black and green), thus demonstrating the robustness of the method.}
\end{figure}

%% file: main.bbl
\begin{thebibliography}{67}
\expandafter\ifx\csname natexlab\endcsname\relax\def\natexlab#1{#1}\fi

\bibitem[{{Abbott} {et~al.}(2022){Abbott}, {Aguena}, {Alarcon}, {Allam}, {Alves}, {Amon}, {Andrade-Oliveira}, {Annis}, {Avila}, {Bacon}, {Baxter}, {Bechtol}, {Becker}, {Bernstein}, {Bhargava}, {Birrer}, {Blazek}, {Brandao-Souza}, {Bridle}, {Brooks}, {Buckley-Geer}, {Burke}, {Camacho}, {Campos}, {Carnero Rosell}, {Carrasco Kind}, {Carretero}, {Castander}, {Cawthon}, {Chang}, {Chen}, {Chen}, {Choi}, {Conselice}, {Cordero}, {Costanzi}, {Crocce}, {da Costa}, {da Silva Pereira}, {Davis}, {Davis}, {De Vicente}, {DeRose}, {Desai}, {Di Valentino}, {Diehl}, {Dietrich}, {Dodelson}, {Doel}, {Doux}, {Drlica-Wagner}, {Eckert}, {Eifler}, {Elsner}, {Elvin-Poole}, {Everett}, {Evrard}, {Fang}, {Farahi}, {Fernandez}, {Ferrero}, {Fert{\'e}}, {Fosalba}, {Friedrich}, {Frieman}, {Garc{\'\i}a-Bellido}, {Gatti}, {Gaztanaga}, {Gerdes}, {Giannantonio}, {Giannini}, {Gruen}, {Gruendl}, {Gschwend}, {Gutierrez}, {Harrison}, {Hartley}, {Herner}, {Hinton}, {Hollowood}, {Honscheid}, {Hoyle}, {Huff}, {Huterer}, {Jain}, {James}, {Jarvis},
  {Jeffrey}, {Jeltema}, {Kovacs}, {Krause}, {Kron}, {Kuehn}, {Kuropatkin}, {Lahav}, {Leget}, {Lemos}, {Liddle}, {Lidman}, {Lima}, {Lin}, {MacCrann}, {Maia}, {Marshall}, {Martini}, {McCullough}, {Melchior}, {Mena-Fern{\'a}ndez}, {Menanteau}, {Miquel}, {Mohr}, {Morgan}, {Muir}, {Myles}, {Nadathur}, {Navarro-Alsina}, {Nichol}, {Ogando}, {Omori}, {Palmese}, {Pandey}, {Park}, {Paz-Chinch{\'o}n}, {Petravick}, {Pieres}, {Plazas Malag{\'o}n}, {Porredon}, {Prat}, {Raveri}, {Rodriguez-Monroy}, {Rollins}, {Romer}, {Roodman}, {Rosenfeld}, {Ross}, {Rykoff}, {Samuroff}, {S{\'a}nchez}, {Sanchez}, {Sanchez}, {Sanchez Cid}, {Scarpine}, {Schubnell}, {Scolnic}, {Secco}, {Serrano}, {Sevilla-Noarbe}, {Sheldon}, {Shin}, {Smith}, {Soares-Santos}, {Suchyta}, {Swanson}, {Tabbutt}, {Tarle}, {Thomas}, {To}, {Troja}, {Troxel}, {Tucker}, {Tutusaus}, {Varga}, {Walker}, {Weaverdyck}, {Wechsler}, {Weller}, {Yanny}, {Yin}, {Zhang}, {Zuntz}, \& {DES Collaboration}}]{2022PhRvD.105b3520A}
{Abbott}, T.~M.~C., {Aguena}, M., {Alarcon}, A., {et~al.} 2022, \href{http://dx.doi.org/10.1103/PhysRevD.105.023520}{\color{blue}\prd}, \href{https://ui.adsabs.harvard.edu/abs/2022PhRvD.105b3520A}{105, 023520}

\bibitem[{{Ade} {et~al.}(2019){Ade}, {Aguirre}, {Ahmed}, {Aiola}, {Ali}, {Alonso}, {Alvarez}, {Arnold}, {Ashton}, {Austermann}, {Awan}, {Baccigalupi}, {Baildon}, {Barron}, {Battaglia}, {Battye}, {Baxter}, {Bazarko}, {Beall}, {Bean}, {Beck}, {Beckman}, {Beringue}, {Bianchini}, {Boada}, {Boettger}, {Bond}, {Borrill}, {Brown}, {Bruno}, {Bryan}, {Calabrese}, {Calafut}, {Calisse}, {Carron}, {Challinor}, {Chesmore}, {Chinone}, {Chluba}, {Cho}, {Choi}, {Coppi}, {Cothard}, {Coughlin}, {Crichton}, {Crowley}, {Crowley}, {Cukierman}, {D'Ewart}, {D{\"u}nner}, {de Haan}, {Devlin}, {Dicker}, {Didier}, {Dobbs}, {Dober}, {Duell}, {Duff}, {Duivenvoorden}, {Dunkley}, {Dusatko}, {Errard}, {Fabbian}, {Feeney}, {Ferraro}, {Flux{\`a}}, {Freese}, {Frisch}, {Frolov}, {Fuller}, {Fuzia}, {Galitzki}, {Gallardo}, {Tomas Galvez Ghersi}, {Gao}, {Gawiser}, {Gerbino}, {Gluscevic}, {Goeckner-Wald}, {Golec}, {Gordon}, {Gralla}, {Green}, {Grigorian}, {Groh}, {Groppi}, {Guan}, {Gudmundsson}, {Han}, {Hargrave}, {Hasegawa}, {Hasselfield},
  {Hattori}, {Haynes}, {Hazumi}, {He}, {Healy}, {Henderson}, {Hervias-Caimapo}, {Hill}, {Hill}, {Hilton}, {Hilton}, {Hincks}, {Hinshaw}, {Hlo{\v{z}}ek}, {Ho}, {Ho}, {Howe}, {Huang}, {Hubmayr}, {Huffenberger}, {Hughes}, {Ijjas}, {Ikape}, {Irwin}, {Jaffe}, {Jain}, {Jeong}, {Kaneko}, {Karpel}, {Katayama}, {Keating}, {Kernasovskiy}, {Keskitalo}, {Kisner}, {Kiuchi}, {Klein}, {Knowles}, {Koopman}, {Kosowsky}, {Krachmalnicoff}, {Kuenstner}, {Kuo}, {Kusaka}, {Lashner}, {Lee}, {Lee}, {Leon}, {Leung}, {Lewis}, {Li}, {Li}, {Limon}, {Linder}, {Lopez-Caraballo}, {Louis}, {Lowry}, {Lungu}, {Madhavacheril}, {Mak}, {Maldonado}, {Mani}, {Mates}, {Matsuda}, {Maurin}, {Mauskopf}, {May}, {McCallum}, {McKenney}, {McMahon}, {Meerburg}, {Meyers}, {Miller}, {Mirmelstein}, {Moodley}, {Munchmeyer}, {Munson}, {Naess}, {Nati}, {Navaroli}, {Newburgh}, {Nguyen}, {Niemack}, {Nishino}, {Orlowski-Scherer}, {Page}, {Partridge}, {Peloton}, {Perrotta}, {Piccirillo}, {Pisano}, {Poletti}, {Puddu}, {Puglisi}, {Raum}, {Reichardt}, {Remazeilles},
  {Rephaeli}, {Riechers}, {Rojas}, {Roy}, {Sadeh}, {Sakurai}, {Salatino}, {Sathyanarayana Rao}, {Schaan}, {Schmittfull}, {Sehgal}, {Seibert}, {Seljak}, {Sherwin}, {Shimon}, {Sierra}, {Sievers}, {Sikhosana}, {Silva-Feaver}, {Simon}, {Sinclair}, {Siritanasak}, {Smith}, {Smith}, {Spergel}, {Staggs}, {Stein}, {Stevens}, {Stompor}, {Suzuki}, {Tajima}, {Takakura}, {Teply}, {Thomas}, {Thorne}, {Thornton}, {Trac}, {Tsai}, {Tucker}, {Ullom}, {Vagnozzi}, {van Engelen}, {Van Lanen}, {Van Winkle}, {Vavagiakis}, {Verg{\`e}s}, {Vissers}, {Wagoner}, {Walker}, {Ward}, {Westbrook}, {Whitehorn}, {Williams}, {Williams}, {Wollack}, {Xu}, {Yu}, {Yu}, {Zago}, {Zhang}, {Zhu}, \& {Simons Observatory Collaboration}}]{2019JCAP...02..056A}
{Ade}, P., {Aguirre}, J., {Ahmed}, Z., {et~al.} 2019, \href{http://dx.doi.org/10.1088/1475-7516/2019/02/056}{\color{blue}\jcap}, \href{https://ui.adsabs.harvard.edu/abs/2019JCAP...02..056A}{2019, 056}

\bibitem[{{Aiola} {et~al.}(2020){Aiola}, {Calabrese}, {Maurin}, {Naess}, {Schmitt}, {Abitbol}, {Addison}, {Ade}, {Alonso}, {Amiri}, {Amodeo}, {Angile}, {Austermann}, {Baildon}, {Battaglia}, {Beall}, {Bean}, {Becker}, {Bond}, {Bruno}, {Calafut}, {Campusano}, {Carrero}, {Chesmore}, {Cho}, {Choi}, {Clark}, {Cothard}, {Crichton}, {Crowley}, {Darwish}, {Datta}, {Denison}, {Devlin}, {Duell}, {Duff}, {Duivenvoorden}, {Dunkley}, {D{\"u}nner}, {Essinger-Hileman}, {Fankhanel}, {Ferraro}, {Fox}, {Fuzia}, {Gallardo}, {Gluscevic}, {Golec}, {Grace}, {Gralla}, {Guan}, {Hall}, {Halpern}, {Han}, {Hargrave}, {Hasselfield}, {Helton}, {Henderson}, {Hensley}, {Hill}, {Hilton}, {Hilton}, {Hincks}, {Hlo{\v{z}}ek}, {Ho}, {Hubmayr}, {Huffenberger}, {Hughes}, {Infante}, {Irwin}, {Jackson}, {Klein}, {Knowles}, {Koopman}, {Kosowsky}, {Lakey}, {Li}, {Li}, {Li}, {Lokken}, {Louis}, {Lungu}, {MacInnis}, {Madhavacheril}, {Maldonado}, {Mallaby-Kay}, {Marsden}, {McMahon}, {Menanteau}, {Moodley}, {Morton}, {Namikawa}, {Nati}, {Newburgh},
  {Nibarger}, {Nicola}, {Niemack}, {Nolta}, {Orlowski-Sherer}, {Page}, {Pappas}, {Partridge}, {Phakathi}, {Pisano}, {Prince}, {Puddu}, {Qu}, {Rivera}, {Robertson}, {Rojas}, {Salatino}, {Schaan}, {Schillaci}, {Sehgal}, {Sherwin}, {Sierra}, {Sievers}, {Sifon}, {Sikhosana}, {Simon}, {Spergel}, {Staggs}, {Stevens}, {Storer}, {Sunder}, {Switzer}, {Thorne}, {Thornton}, {Trac}, {Treu}, {Tucker}, {Vale}, {Van Engelen}, {Van Lanen}, {Vavagiakis}, {Wagoner}, {Wang}, {Ward}, {Wollack}, {Xu}, {Zago}, \& {Zhu}}]{2020JCAP...12..047A}
{Aiola}, S., {Calabrese}, E., {Maurin}, L., {et~al.} 2020, \href{http://dx.doi.org/10.1088/1475-7516/2020/12/047}{\color{blue}\jcap}, \href{https://ui.adsabs.harvard.edu/abs/2020JCAP...12..047A}{2020, 047}

\bibitem[{{Alam} {et~al.}(2015){Alam}, {Albareti}, {Allende Prieto}, {Anders}, {Anderson}, {Anderton}, {Andrews}, {Armengaud}, {Aubourg}, {Bailey}, {Basu}, {Bautista}, {Beaton}, {Beers}, {Bender}, {Berlind}, {Beutler}, {Bhardwaj}, {Bird}, {Bizyaev}, {Blake}, {Blanton}, {Blomqvist}, {Bochanski}, {Bolton}, {Bovy}, {Shelden Bradley}, {Brandt}, {Brauer}, {Brinkmann}, {Brown}, {Brownstein}, {Burden}, {Burtin}, {Busca}, {Cai}, {Capozzi}, {Carnero Rosell}, {Carr}, {Carrera}, {Chambers}, {Chaplin}, {Chen}, {Chiappini}, {Chojnowski}, {Chuang}, {Clerc}, {Comparat}, {Covey}, {Croft}, {Cuesta}, {Cunha}, {da Costa}, {Da Rio}, {Davenport}, {Dawson}, {De Lee}, {Delubac}, {Deshpande}, {Dhital}, {Dutra-Ferreira}, {Dwelly}, {Ealet}, {Ebelke}, {Edmondson}, {Eisenstein}, {Ellsworth}, {Elsworth}, {Epstein}, {Eracleous}, {Escoffier}, {Esposito}, {Evans}, {Fan}, {Fern{\'a}ndez-Alvar}, {Feuillet}, {Filiz Ak}, {Finley}, {Finoguenov}, {Flaherty}, {Fleming}, {Font-Ribera}, {Foster}, {Frinchaboy}, {Galbraith-Frew}, {Garc{\'\i}a},
  {Garc{\'\i}a-Hern{\'a}ndez}, {Garc{\'\i}a P{\'e}rez}, {Gaulme}, {Ge}, {G{\'e}nova-Santos}, {Georgakakis}, {Ghezzi}, {Gillespie}, {Girardi}, {Goddard}, {Gontcho}, {Gonz{\'a}lez Hern{\'a}ndez}, {Grebel}, {Green}, {Grieb}, {Grieves}, {Gunn}, {Guo}, {Harding}, {Hasselquist}, {Hawley}, {Hayden}, {Hearty}, {Hekker}, {Ho}, {Hogg}, {Holley-Bockelmann}, {Holtzman}, {Honscheid}, {Huber}, {Huehnerhoff}, {Ivans}, {Jiang}, {Johnson}, {Kinemuchi}, {Kirkby}, {Kitaura}, {Klaene}, {Knapp}, {Kneib}, {Koenig}, {Lam}, {Lan}, {Lang}, {Laurent}, {Le Goff}, {Leauthaud}, {Lee}, {Lee}, {Licquia}, {Liu}, {Long}, {L{\'o}pez-Corredoira}, {Lorenzo-Oliveira}, {Lucatello}, {Lundgren}, {Lupton}, {Mack}, {Mahadevan}, {Maia}, {Majewski}, {Malanushenko}, {Malanushenko}, {Manchado}, {Manera}, {Mao}, {Maraston}, {Marchwinski}, {Margala}, {Martell}, {Martig}, {Masters}, {Mathur}, {McBride}, {McGehee}, {McGreer}, {McMahon}, {M{\'e}nard}, {Menzel}, {Merloni}, {M{\'e}sz{\'a}ros}, {Miller}, {Miralda-Escud{\'e}}, {Miyatake}, {Montero-Dorta}, {More},
  {Morganson}, {Morice-Atkinson}, {Morrison}, {Mosser}, {Muna}, {Myers}, {Nandra}, {Newman}, {Neyrinck}, {Nguyen}, {Nichol}, {Nidever}, {Noterdaeme}, {Nuza}, {O'Connell}, {O'Connell}, {O'Connell}, {Ogando}, {Olmstead}, {Oravetz}, {Oravetz}, {Osumi}, {Owen}, {Padgett}, {Padmanabhan}, {Paegert}, {Palanque-Delabrouille}, {Pan}, {Parejko}, {P{\^a}ris}, {Park}, {Pattarakijwanich}, {Pellejero-Ibanez}, {Pepper}, {Percival}, {P{\'e}rez-Fournon}, {P{\'e}rez-R{\`a}fols}, {Petitjean}, {Pieri}, {Pinsonneault}, {Porto de Mello}, {Prada}, {Prakash}, {Price-Whelan}, {Protopapas}, {Raddick}, {Rahman}, {Reid}, {Rich}, {Rix}, {Robin}, {Rockosi}, {Rodrigues}, {Rodr{\'\i}guez-Torres}, {Roe}, {Ross}, {Ross}, {Rossi}, {Ruan}, {Rubi{\~n}o-Mart{\'\i}n}, {Rykoff}, {Salazar-Albornoz}, {Salvato}, {Samushia}, {S{\'a}nchez}, {Santiago}, {Sayres}, {Schiavon}, {Schlegel}, {Schmidt}, {Schneider}, {Schultheis}, {Schwope}, {Sc{\'o}ccola}, {Scott}, {Sellgren}, {Seo}, {Serenelli}, {Shane}, {Shen}, {Shetrone}, {Shu}, {Silva Aguirre}, {Sivarani},
  {Skrutskie}, {Slosar}, {Smith}, {Sobreira}, {Souto}, {Stassun}, {Steinmetz}, {Stello}, {Strauss}, {Streblyanska}, {Suzuki}, {Swanson}, {Tan}, {Tayar}, {Terrien}, {Thakar}, {Thomas}, {Thomas}, {Thompson}, {Tinker}, {Tojeiro}, {Troup}, {Vargas-Maga{\~n}a}, {Vazquez}, {Verde}, {Viel}, {Vogt}, {Wake}, {Wang}, {Weaver}, {Weinberg}, {Weiner}, {White}, {Wilson}, {Wisniewski}, {Wood-Vasey}, {Ye`che}, {York}, {Zakamska}, {Zamora}, {Zasowski}, {Zehavi}, {Zhao}, {Zheng}, {Zhou}, {Zhou}, {Zou}, \& {Zhu}}]{2015ApJS..219...12A}
{Alam}, S., {Albareti}, F.~D., {Allende Prieto}, C., {et~al.} 2015, \href{http://dx.doi.org/10.1088/0067-0049/219/1/12}{\color{blue}\apjs}, \href{https://ui.adsabs.harvard.edu/abs/2015ApJS..219...12A}{219, 12}

\bibitem[{{Alam} {et~al.}(2021){Alam}, {Aubert}, {Avila}, {Balland}, {Bautista}, {Bershady}, {Bizyaev}, {Blanton}, {Bolton}, {Bovy}, {Brinkmann}, {Brownstein}, {Burtin}, {Chabanier}, {Chapman}, {Choi}, {Chuang}, {Comparat}, {Cousinou}, {Cuceu}, {Dawson}, {de la Torre}, {de Mattia}, {Agathe}, {des Bourboux}, {Escoffier}, {Etourneau}, {Farr}, {Font-Ribera}, {Frinchaboy}, {Fromenteau}, {Gil-Mar{\'\i}n}, {Le Goff}, {Gonzalez-Morales}, {Gonzalez-Perez}, {Grabowski}, {Guy}, {Hawken}, {Hou}, {Kong}, {Parker}, {Klaene}, {Kneib}, {Lin}, {Long}, {Lyke}, {de la Macorra}, {Martini}, {Masters}, {Mohammad}, {Moon}, {Mueller}, {Mu{\~n}oz-Guti{\'e}rrez}, {Myers}, {Nadathur}, {Neveux}, {Newman}, {Noterdaeme}, {Oravetz}, {Oravetz}, {Palanque-Delabrouille}, {Pan}, {Paviot}, {Percival}, {P{\'e}rez-R{\`a}fols}, {Petitjean}, {Pieri}, {Prakash}, {Raichoor}, {Ravoux}, {Rezaie}, {Rich}, {Ross}, {Rossi}, {Ruggeri}, {Ruhlmann-Kleider}, {S{\'a}nchez}, {S{\'a}nchez}, {S{\'a}nchez-Gallego}, {Sayres}, {Schneider}, {Seo}, {Shafieloo},
  {Slosar}, {Smith}, {Stermer}, {Tamone}, {Tinker}, {Tojeiro}, {Vargas-Maga{\~n}a}, {Variu}, {Wang}, {Weaver}, {Weijmans}, {Y{\`e}che}, {Zarrouk}, {Zhao}, {Zhao}, \& {Zheng}}]{2021PhRvD.103h3533A}
{Alam}, S., {Aubert}, M., {Avila}, S., {et~al.} 2021, \href{http://dx.doi.org/10.1103/PhysRevD.103.083533}{\color{blue}\prd}, \href{https://ui.adsabs.harvard.edu/abs/2021PhRvD.103h3533A}{103, 083533}

\bibitem[{{Alonso} {et~al.}(2019){Alonso}, {Sanchez}, {Slosar}, \& {LSST Dark Energy Science Collaboration}}]{1809.09603}
{Alonso}, D., {Sanchez}, J., {Slosar}, A., \& {LSST Dark Energy Science Collaboration}. 2019, \href{http://dx.doi.org/10.1093/mnras/stz093}{\color{blue}\mnras}, \href{https://ui.adsabs.harvard.edu/abs/2019MNRAS.484.4127A}{484, 4127}

\bibitem[{{Alsing} \& {Handley}(2021)}]{2021MNRAS.505L..95A}
{Alsing}, J. \& {Handley}, W. 2021, \href{http://dx.doi.org/10.1093/mnrasl/slab057}{\color{blue}\mnras}, \href{https://ui.adsabs.harvard.edu/abs/2021MNRAS.505L..95A}{505, L95}

\bibitem[{{Alsing} {et~al.}(2018){Alsing}, {Wandelt}, \& {Feeney}}]{2018MNRAS.477.2874A}
{Alsing}, J., {Wandelt}, B., \& {Feeney}, S. 2018, \href{http://dx.doi.org/10.1093/mnras/sty819}{\color{blue}\mnras}, \href{https://ui.adsabs.harvard.edu/abs/2018MNRAS.477.2874A}{477, 2874}

\bibitem[{{Amendola} {et~al.}(2018){Amendola}, {Appleby}, {Avgoustidis}, {Bacon}, {Baker}, {Baldi}, {Bartolo}, {Blanchard}, {Bonvin}, {Borgani}, {Branchini}, {Burrage}, {Camera}, {Carbone}, {Casarini}, {Cropper}, {de Rham}, {Dietrich}, {Di Porto}, {Durrer}, {Ealet}, {Ferreira}, {Finelli}, {Garc{\'\i}a-Bellido}, {Giannantonio}, {Guzzo}, {Heavens}, {Heisenberg}, {Heymans}, {Hoekstra}, {Hollenstein}, {Holmes}, {Hwang}, {Jahnke}, {Kitching}, {Koivisto}, {Kunz}, {La Vacca}, {Linder}, {March}, {Marra}, {Martins}, {Majerotto}, {Markovic}, {Marsh}, {Marulli}, {Massey}, {Mellier}, {Montanari}, {Mota}, {Nunes}, {Percival}, {Pettorino}, {Porciani}, {Quercellini}, {Read}, {Rinaldi}, {Sapone}, {Sawicki}, {Scaramella}, {Skordis}, {Simpson}, {Taylor}, {Thomas}, {Trotta}, {Verde}, {Vernizzi}, {Vollmer}, {Wang}, {Weller}, \& {Zlosnik}}]{2018LRR....21....2A}
{Amendola}, L., {Appleby}, S., {Avgoustidis}, A., {et~al.} 2018, \href{http://dx.doi.org/10.1007/s41114-017-0010-3}{\color{blue}Living Reviews in Relativity}, \href{https://ui.adsabs.harvard.edu/abs/2018LRR....21....2A}{21, 2}

\bibitem[{{Aric{\`o}} {et~al.}(2021){Aric{\`o}}, {Angulo}, {Contreras}, {Ondaro-Mallea}, {Pellejero-Iba{\~n}ez}, \& {Zennaro}}]{2021MNRAS.506.4070A}
{Aric{\`o}}, G., {Angulo}, R.~E., {Contreras}, S., {et~al.} 2021, \href{http://dx.doi.org/10.1093/mnras/stab1911}{\color{blue}\mnras}, \href{https://ui.adsabs.harvard.edu/abs/2021MNRAS.506.4070A}{506, 4070}

\bibitem[{{Asgari} {et~al.}(2021){Asgari}, {Lin}, {Joachimi}, {Giblin}, {Heymans}, {Hildebrandt}, {Kannawadi}, {St{\"o}lzner}, {Tr{\"o}ster}, {van den Busch}, {Wright}, {Bilicki}, {Blake}, {de Jong}, {Dvornik}, {Erben}, {Getman}, {Hoekstra}, {K{\"o}hlinger}, {Kuijken}, {Miller}, {Radovich}, {Schneider}, {Shan}, \& {Valentijn}}]{2021A&A...645A.104A}
{Asgari}, M., {Lin}, C.-A., {Joachimi}, B., {et~al.} 2021, \href{http://dx.doi.org/10.1051/0004-6361/202039070}{\color{blue}\aap}, \href{https://ui.adsabs.harvard.edu/abs/2021A&A...645A.104A}{645, A104}

\bibitem[{{Bartlett} {et~al.}(2024{\natexlab{a}}){Bartlett}, {Kammerer}, {Kronberger}, {Desmond}, {Ferreira}, {Wandelt}, {Burlacu}, {Alonso}, \& {Zennaro}}]{2023arXiv231115865B}
{Bartlett}, D.~J., {Kammerer}, L., {Kronberger}, G., {et~al.} 2024{\natexlab{a}}, \href{http://dx.doi.org/10.1051/0004-6361/202348811}{\color{blue}\aap}, \href{https://ui.adsabs.harvard.edu/abs/2024A&A...686A.209B}{686, A209}

\bibitem[{{Bartlett} {et~al.}(2024{\natexlab{b}}){Bartlett}, {Wandelt}, {Zennaro}, {Ferreira}, \& {Desmond}}]{2024A&A...686A.150B}
{Bartlett}, D.~J., {Wandelt}, B.~D., {Zennaro}, M., {Ferreira}, P.~G., \& {Desmond}, H. 2024{\natexlab{b}}, \href{http://dx.doi.org/10.1051/0004-6361/202449854}{\color{blue}\aap}, \href{https://ui.adsabs.harvard.edu/abs/2024A&A...686A.150B}{686, A150}

\bibitem[{{Bevins} {et~al.}(2023){Bevins}, {Handley}, {Lemos}, {Sims}, {de Lera Acedo}, {Fialkov}, \& {Alsing}}]{2023MNRAS.526.4613B}
{Bevins}, H. T.~J., {Handley}, W.~J., {Lemos}, P., {et~al.} 2023, \href{http://dx.doi.org/10.1093/mnras/stad2997}{\color{blue}\mnras}, \href{https://ui.adsabs.harvard.edu/abs/2023MNRAS.526.4613B}{526, 4613}

\bibitem[{{Bi{\'n}kowski} {et~al.}(2018){Bi{\'n}kowski}, {Sutherland}, {Arbel}, \& {Gretton}}]{2018arXiv180101401B}
{Bi{\'n}kowski}, M., {Sutherland}, D.~J., {Arbel}, M., \& {Gretton}, A. 2018, \href{https://ui.adsabs.harvard.edu/abs/2018arXiv180101401B}{\href{http://dx.doi.org/10.48550/arXiv.1801.01401}{\color{blue}arXiv e-prints}, arXiv:1801.01401}

\bibitem[{{Blake} {et~al.}(2016){Blake}, {Amon}, {Childress}, {Erben}, {Glazebrook}, {Harnois-Deraps}, {Heymans}, {Hildebrandt}, {Hinton}, {Janssens}, {Johnson}, {Joudaki}, {Klaes}, {Kuijken}, {Lidman}, {Marin}, {Parkinson}, {Poole}, \& {Wolf}}]{2016MNRAS.462.4240B}
{Blake}, C., {Amon}, A., {Childress}, M., {et~al.} 2016, \href{http://dx.doi.org/10.1093/mnras/stw1990}{\color{blue}\mnras}, \href{https://ui.adsabs.harvard.edu/abs/2016MNRAS.462.4240B}{462, 4240}

\bibitem[{{Blas} {et~al.}(2011){Blas}, {Lesgourgues}, \& {Tram}}]{1104.2933}
{Blas}, D., {Lesgourgues}, J., \& {Tram}, T. 2011, \href{http://dx.doi.org/10.1088/1475-7516/2011/07/034}{\color{blue}\jcap}, \href{https://ui.adsabs.harvard.edu/abs/2011JCAP...07..034B}{2011, 034}

\bibitem[{{Campagne} {et~al.}(2023){Campagne}, {Lanusse}, {Zuntz}, {Boucaud}, {Casas}, {Karamanis}, {Kirkby}, {Lanzieri}, {Peel}, \& {Li}}]{2023OJAp....6E..15C}
{Campagne}, J.-E., {Lanusse}, F., {Zuntz}, J., {et~al.} 2023, \href{http://dx.doi.org/10.21105/astro.2302.05163}{\color{blue}The Open Journal of Astrophysics}, \href{https://ui.adsabs.harvard.edu/abs/2023OJAp....6E..15C}{6, 15}

\bibitem[{{Chisari} {et~al.}(2019){Chisari}, {Alonso}, {Krause}, {Leonard}, {Bull}, {Neveu}, {Villarreal}, {Singh}, {McClintock}, {Ellison}, {Du}, {Zuntz}, {Mead}, {Joudaki}, {Lorenz}, {Tr{\"o}ster}, {Sanchez}, {Lanusse}, {Ishak}, {Hlozek}, {Blazek}, {Campagne}, {Almoubayyed}, {Eifler}, {Kirby}, {Kirkby}, {Plaszczynski}, {Slosar}, {Vrastil}, {Wagoner}, \& {LSST Dark Energy Science Collaboration}}]{1812.05995}
{Chisari}, N.~E., {Alonso}, D., {Krause}, E., {et~al.} 2019, \href{http://dx.doi.org/10.3847/1538-4365/ab1658}{\color{blue}\apjs}, \href{https://ui.adsabs.harvard.edu/abs/2019ApJS..242....2C}{242, 2}

\bibitem[{{Dark Energy Survey and Kilo-Degree Survey Collaboration} {et~al.}(2023){Dark Energy Survey and Kilo-Degree Survey Collaboration}, {Abbott}, {Aguena}, {Alarcon}, {Alves}, {Amon}, {Andrade-Oliveira}, {Asgari}, {Avila}, {Bacon}, {Bechtol}, {Becker}, {Bernstein}, {Bertin}, {Bilicki}, {Blazek}, {Bocquet}, {Brooks}, {Burger}, {Burke}, {Camacho}, {Campos}, {Carnero Rosell}, {Carrasco Kind}, {Carretero}, {Castander}, {Cawthon}, {Chang}, {Chen}, {Choi}, {Conselice}, {Cordero}, {Crocce}, {da Costa}, {da Silva Pereira}, {Dalal}, {Davis}, {de Jong}, {DeRose}, {Desai}, {Diehl}, {Dodelson}, {Doel}, {Doux}, {Drlica-Wagner}, {Dvornik}, {Eckert}, {Eifler}, {Elvin-Poole}, {Everett}, {Fang}, {Ferrero}, {Fert{\'e}}, {Flaugher}, {Friedrich}, {Frieman}, {Garc{\'\i}a-Bellido}, {Gatti}, {Giannini}, {Giblin}, {Gruen}, {Gruendl}, {Gutierrez}, {Harrison}, {Hartley}, {Herner}, {Heymans}, {Hildebrandt}, {Hinton}, {Hoekstra}, {Hollowood}, {Honscheid}, {Huang}, {Huff}, {Huterer}, {James}, {Jarvis}, {Jeffrey}, {Jeltema},
  {Joachimi}, {Joudaki}, {Kannawadi}, {Krause}, {Kuehn}, {Kuijken}, {Kuropatkin}, {Lahav}, {Leget}, {Lemos}, {Li}, {Li}, {Liddle}, {Lima}, {Lin}, {Lin}, {MacCrann}, {Mahony}, {Marshall}, {McCullough}, {Mena-Fern{\'a}ndez}, {Menanteau}, {Miquel}, {Mohr}, {Muir}, {Myles}, {Napolitano}, {Navarro-Alsina}, {Ogando}, {Palmese}, {Pandey}, {Park}, {Paterno}, {Peacock}, {Petravick}, {Pieres}, {Plazas Malag{\'o}n}, {Porredon}, {Prat}, {Radovich}, {Raveri}, {Reischke}, {Robertson}, {Rollins}, {Romer}, {Roodman}, {Rykoff}, {Samuroff}, {S{\'a}nchez}, {Sanchez}, {Sanchez}, {Schneider}, {Secco}, {Sevilla-Noarbe}, {Shan}, {Sheldon}, {Shin}, {Sif{\'o}n}, {Smith}, {Soares-Santos}, {St{\"o}lzner}, {Suchyta}, {Swanson}, {Tarle}, {Thomas}, {To}, {Troxel}, {Tr{\"o}ster}, {Tutusaus}, {van den Busch}, {Varga}, {Walker}, {Weaverdyck}, {Wechsler}, {Weller}, {Wiseman}, {Wright}, {Yanny}, {Yin}, {Yoon}, {Zhang}, \& {Zuntz}}]{2023OJAp....6E..36D}
{Dark Energy Survey and Kilo-Degree Survey Collaboration}, {Abbott}, T.~M.~C., {Aguena}, M., {et~al.} 2023, \href{http://dx.doi.org/10.21105/astro.2305.17173}{\color{blue}The Open Journal of Astrophysics}, \href{https://ui.adsabs.harvard.edu/abs/2023OJAp....6E..36D}{6, 36}

\bibitem[{{DESI Collaboration} {et~al.}(2024){DESI Collaboration}, {Adame}, {Aguilar}, {Ahlen}, {Alam}, {Alexander}, {Alvarez}, {Alves}, {Anand}, {Andrade}, {Armengaud}, {Avila}, {Aviles}, {Awan}, {Bahr-Kalus}, {Bailey}, {Baltay}, {Bault}, {Behera}, {BenZvi}, {Bera}, {Beutler}, {Bianchi}, {Blake}, {Blum}, {Brieden}, {Brodzeller}, {Brooks}, {Buckley-Geer}, {Burtin}, {Calderon}, {Canning}, {Carnero Rosell}, {Cereskaite}, {Cervantes-Cota}, {Chabanier}, {Chaussidon}, {Chaves-Montero}, {Chen}, {Chen}, {Claybaugh}, {Cole}, {Cuceu}, {Davis}, {Dawson}, {de la Macorra}, {de Mattia}, {Deiosso}, {Dey}, {Dey}, {Ding}, {Doel}, {Edelstein}, {Eftekharzadeh}, {Eisenstein}, {Elliott}, {Fagrelius}, {Fanning}, {Ferraro}, {Ereza}, {Findlay}, {Flaugher}, {Font-Ribera}, {Forero-S{\'a}nchez}, {Forero-Romero}, {Frenk}, {Garcia-Quintero}, {Gazta{\~n}aga}, {Gil-Mar{\'\i}n}, {Gontcho}, {Gonzalez-Morales}, {Gonzalez-Perez}, {Gordon}, {Green}, {Gruen}, {Gsponer}, {Gutierrez}, {Guy}, {Hadzhiyska}, {Hahn}, {Hanif}, {Herrera-Alcantar},
  {Honscheid}, {Howlett}, {Huterer}, {Ir{\v{s}}i{\v{c}}}, {Ishak}, {Juneau}, {Kara{\c{c}}ayl{\i}}, {Kehoe}, {Kent}, {Kirkby}, {Kremin}, {Krolewski}, {Lai}, {Lan}, {Landriau}, {Lang}, {Lasker}, {Le Goff}, {Le Guillou}, {Leauthaud}, {Levi}, {Li}, {Linder}, {Lodha}, {Magneville}, {Manera}, {Margala}, {Martini}, {Maus}, {McDonald}, {Medina-Varela}, {Meisner}, {Mena-Fern{\'a}ndez}, {Miquel}, {Moon}, {Moore}, {Moustakas}, {Mudur}, {Mueller}, {Mu{\~n}oz-Guti{\'e}rrez}, {Myers}, {Nadathur}, {Napolitano}, {Neveux}, {Newman}, {Nguyen}, {Nie}, {Niz}, {Noriega}, {Padmanabhan}, {Paillas}, {Palanque-Delabrouille}, {Pan}, {Penmetsa}, {Percival}, {Pieri}, {Pinon}, {Poppett}, {Porredon}, {Prada}, {P{\'e}rez-Fern{\'a}ndez}, {P{\'e}rez-R{\`a}fols}, {Rabinowitz}, {Raichoor}, {Ram{\'\i}rez-P{\'e}rez}, {Ramirez-Solano}, {Ravoux}, {Rashkovetskyi}, {Rezaie}, {Rich}, {Rocher}, {Rockosi}, {Roe}, {Rosado-Marin}, {Ross}, {Rossi}, {Ruggeri}, {Ruhlmann-Kleider}, {Samushia}, {Sanchez}, {Saulder}, {Schlafly}, {Schlegel}, {Schubnell}, {Seo},
  {Shafieloo}, {Sharples}, {Silber}, {Slosar}, {Smith}, {Sprayberry}, {Tan}, {Tarl{\'e}}, {Taylor}, {Trusov}, {Ure{\~n}a-L{\'o}pez}, {Vaisakh}, {Valcin}, {Valdes}, {Vargas-Maga{\~n}a}, {Verde}, {Walther}, {Wang}, {Wang}, {Weaver}, {Weaverdyck}, {Wechsler}, {Weinberg}, {White}, {Yu}, {Yu}, {Yuan}, {Y{\`e}che}, {Zaborowski}, {Zarrouk}, {Zhang}, {Zhao}, {Zhao}, {Zhou}, {Zhuang}, \& {Zou}}]{2024arXiv240403002D}
{DESI Collaboration}, {Adame}, A.~G., {Aguilar}, J., {et~al.} 2024, \href{https://ui.adsabs.harvard.edu/abs/2024arXiv240403002D}{\href{http://dx.doi.org/10.48550/arXiv.2404.03002}{\color{blue}arXiv e-prints}, arXiv:2404.03002}

\bibitem[{{Feroz} {et~al.}(2019){Feroz}, {Hobson}, {Cameron}, \& {Pettitt}}]{2019OJAp....2E..10F}
{Feroz}, F., {Hobson}, M.~P., {Cameron}, E., \& {Pettitt}, A.~N. 2019, \href{http://dx.doi.org/10.21105/astro.1306.2144}{\color{blue}The Open Journal of Astrophysics}, \href{https://ui.adsabs.harvard.edu/abs/2019OJAp....2E..10F}{2, 10}

\bibitem[{{Foreman-Mackey} {et~al.}(2013){Foreman-Mackey}, {Hogg}, {Lang}, \& {Goodman}}]{2013PASP..125..306F}
{Foreman-Mackey}, D., {Hogg}, D.~W., {Lang}, D., \& {Goodman}, J. 2013, \href{http://dx.doi.org/10.1086/670067}{\color{blue}\pasp}, \href{https://ui.adsabs.harvard.edu/abs/2013PASP..125..306F}{125, 306}

\bibitem[{{Garc{\'\i}a-Garc{\'\i}a} {et~al.}(2021){Garc{\'\i}a-Garc{\'\i}a}, {Ruiz-Zapatero}, {Alonso}, {Bellini}, {Ferreira}, {Mueller}, {Nicola}, \& {Ruiz-Lapuente}}]{2021JCAP...10..030G}
{Garc{\'\i}a-Garc{\'\i}a}, C., {Ruiz-Zapatero}, J., {Alonso}, D., {et~al.} 2021, \href{http://dx.doi.org/10.1088/1475-7516/2021/10/030}{\color{blue}\jcap}, \href{https://ui.adsabs.harvard.edu/abs/2021JCAP...10..030G}{2021, 030}

\bibitem[{{Garc{\'\i}a-Garc{\'\i}a} {et~al.}(2024){Garc{\'\i}a-Garc{\'\i}a}, {Zennaro}, {Aric{\`o}}, {Alonso}, \& {Angulo}}]{2024arXiv240313794G}
{Garc{\'\i}a-Garc{\'\i}a}, C., {Zennaro}, M., {Aric{\`o}}, G., {Alonso}, D., \& {Angulo}, R.~E. 2024, \href{http://dx.doi.org/10.1088/1475-7516/2024/08/024}{\color{blue}\jcap}, \href{https://ui.adsabs.harvard.edu/abs/2024JCAP...08..024G}{2024, 024}

\bibitem[{{Hadzhiyska} {et~al.}(2020){Hadzhiyska}, {Alonso}, {Nicola}, \& {Slosar}}]{2007.14989}
{Hadzhiyska}, B., {Alonso}, D., {Nicola}, A., \& {Slosar}, A. 2020, \href{http://dx.doi.org/10.1088/1475-7516/2020/10/056}{\color{blue}\jcap}, \href{https://ui.adsabs.harvard.edu/abs/2020JCAP...10..056H}{2020, 056}

\bibitem[{{Hang} {et~al.}(2021){Hang}, {Alam}, {Peacock}, \& {Cai}}]{2010.00466}
{Hang}, Q., {Alam}, S., {Peacock}, J.~A., \& {Cai}, Y.-C. 2021, \href{http://dx.doi.org/10.1093/mnras/staa3738}{\color{blue}\mnras}, \href{https://ui.adsabs.harvard.edu/abs/2021MNRAS.501.1481H}{501, 1481}

\bibitem[{{Harris} {et~al.}(2020){Harris}, {Millman}, {van der Walt}, {Gommers}, {Virtanen}, {Cournapeau}, {Wieser}, {Taylor}, {Berg}, {Smith}, {Kern}, {Picus}, {Hoyer}, {van Kerkwijk}, {Brett}, {Haldane}, {del R{\'\i}o}, {Wiebe}, {Peterson}, {G{\'e}rard-Marchant}, {Sheppard}, {Reddy}, {Weckesser}, {Abbasi}, {Gohlke}, \& {Oliphant}}]{harris2020array}
{Harris}, C.~R., {Millman}, K.~J., {van der Walt}, S.~J., {et~al.} 2020, \href{http://dx.doi.org/10.1038/s41586-020-2649-2}{\color{blue}\nat}, \href{https://ui.adsabs.harvard.edu/abs/2020Natur.585..357H}{585, 357}

\bibitem[{{Heavens} {et~al.}(2017{\natexlab{a}}){Heavens}, {Fantaye}, {Mootoovaloo}, {Eggers}, {Hosenie}, {Kroon}, \& {Sellentin}}]{2017arXiv170403472H}
{Heavens}, A., {Fantaye}, Y., {Mootoovaloo}, A., {et~al.} 2017{\natexlab{a}}, \href{https://ui.adsabs.harvard.edu/abs/2017arXiv170403472H}{\href{http://dx.doi.org/10.48550/arXiv.1704.03472}{\color{blue}arXiv e-prints}, arXiv:1704.03472}

\bibitem[{{Heavens} {et~al.}(2017{\natexlab{b}}){Heavens}, {Fantaye}, {Sellentin}, {Eggers}, {Hosenie}, {Kroon}, \& {Mootoovaloo}}]{2017PhRvL.119j1301H}
{Heavens}, A., {Fantaye}, Y., {Sellentin}, E., {et~al.} 2017{\natexlab{b}}, \href{http://dx.doi.org/10.1103/PhysRevLett.119.101301}{\color{blue}\prl}, \href{https://ui.adsabs.harvard.edu/abs/2017PhRvL.119j1301H}{119, 101301}

\bibitem[{{Heymans} {et~al.}(2021){Heymans}, {Tr{\"o}ster}, {Asgari}, {Blake}, {Hildebrandt}, {Joachimi}, {Kuijken}, {Lin}, {S{\'a}nchez}, {van den Busch}, {Wright}, {Amon}, {Bilicki}, {de Jong}, {Crocce}, {Dvornik}, {Erben}, {Fortuna}, {Getman}, {Giblin}, {Glazebrook}, {Hoekstra}, {Joudaki}, {Kannawadi}, {K{\"o}hlinger}, {Lidman}, {Miller}, {Napolitano}, {Parkinson}, {Schneider}, {Shan}, {Valentijn}, {Verdoes Kleijn}, \& {Wolf}}]{2021A&A...646A.140H}
{Heymans}, C., {Tr{\"o}ster}, T., {Asgari}, M., {et~al.} 2021, \href{http://dx.doi.org/10.1051/0004-6361/202039063}{\color{blue}\aap}, \href{https://ui.adsabs.harvard.edu/abs/2021A&A...646A.140H}{646, A140}

\bibitem[{{Hou} {et~al.}(2021){Hou}, {S{\'a}nchez}, {Ross}, {Smith}, {Neveux}, {Bautista}, {Burtin}, {Zhao}, {Scoccimarro}, {Dawson}, {de Mattia}, {de la Macorra}, {du Mas des Bourboux}, {Eisenstein}, {Gil-Mar{\'\i}n}, {Lyke}, {Mohammad}, {Mueller}, {Percival}, {Rossi}, {Vargas Maga{\~n}a}, {Zarrouk}, {Zhao}, {Brinkmann}, {Brownstein}, {Chuang}, {Myers}, {Newman}, {Schneider}, \& {Vivek}}]{2007.08998}
{Hou}, J., {S{\'a}nchez}, A.~G., {Ross}, A.~J., {et~al.} 2021, \href{http://dx.doi.org/10.1093/mnras/staa3234}{\color{blue}\mnras}, \href{https://ui.adsabs.harvard.edu/abs/2021MNRAS.500.1201H}{500, 1201}

\bibitem[{{Ivezi{\'c}} {et~al.}(2019){Ivezi{\'c}}, {Kahn}, {Tyson}, {Abel}, {Acosta}, {Allsman}, {Alonso}, {AlSayyad}, {Anderson}, {Andrew}, {Angel}, {Angeli}, {Ansari}, {Antilogus}, {Araujo}, {Armstrong}, {Arndt}, {Astier}, {Aubourg}, {Auza}, {Axelrod}, {Bard}, {Barr}, {Barrau}, {Bartlett}, {Bauer}, {Bauman}, {Baumont}, {Bechtol}, {Bechtol}, {Becker}, {Becla}, {Beldica}, {Bellavia}, {Bianco}, {Biswas}, {Blanc}, {Blazek}, {Blandford}, {Bloom}, {Bogart}, {Bond}, {Booth}, {Borgland}, {Borne}, {Bosch}, {Boutigny}, {Brackett}, {Bradshaw}, {Brandt}, {Brown}, {Bullock}, {Burchat}, {Burke}, {Cagnoli}, {Calabrese}, {Callahan}, {Callen}, {Carlin}, {Carlson}, {Chandrasekharan}, {Charles-Emerson}, {Chesley}, {Cheu}, {Chiang}, {Chiang}, {Chirino}, {Chow}, {Ciardi}, {Claver}, {Cohen-Tanugi}, {Cockrum}, {Coles}, {Connolly}, {Cook}, {Cooray}, {Covey}, {Cribbs}, {Cui}, {Cutri}, {Daly}, {Daniel}, {Daruich}, {Daubard}, {Daues}, {Dawson}, {Delgado}, {Dellapenna}, {de Peyster}, {de Val-Borro}, {Digel}, {Doherty}, {Dubois},
  {Dubois-Felsmann}, {Durech}, {Economou}, {Eifler}, {Eracleous}, {Emmons}, {Fausti Neto}, {Ferguson}, {Figueroa}, {Fisher-Levine}, {Focke}, {Foss}, {Frank}, {Freemon}, {Gangler}, {Gawiser}, {Geary}, {Gee}, {Geha}, {Gessner}, {Gibson}, {Gilmore}, {Glanzman}, {Glick}, {Goldina}, {Goldstein}, {Goodenow}, {Graham}, {Gressler}, {Gris}, {Guy}, {Guyonnet}, {Haller}, {Harris}, {Hascall}, {Haupt}, {Hernandez}, {Herrmann}, {Hileman}, {Hoblitt}, {Hodgson}, {Hogan}, {Howard}, {Huang}, {Huffer}, {Ingraham}, {Innes}, {Jacoby}, {Jain}, {Jammes}, {Jee}, {Jenness}, {Jernigan}, {Jevremovi{\'c}}, {Johns}, {Johnson}, {Johnson}, {Jones}, {Juramy-Gilles}, {Juri{\'c}}, {Kalirai}, {Kallivayalil}, {Kalmbach}, {Kantor}, {Karst}, {Kasliwal}, {Kelly}, {Kessler}, {Kinnison}, {Kirkby}, {Knox}, {Kotov}, {Krabbendam}, {Krughoff}, {Kub{\'a}nek}, {Kuczewski}, {Kulkarni}, {Ku}, {Kurita}, {Lage}, {Lambert}, {Lange}, {Langton}, {Le Guillou}, {Levine}, {Liang}, {Lim}, {Lintott}, {Long}, {Lopez}, {Lotz}, {Lupton}, {Lust}, {MacArthur}, {Mahabal},
  {Mandelbaum}, {Markiewicz}, {Marsh}, {Marshall}, {Marshall}, {May}, {McKercher}, {McQueen}, {Meyers}, {Migliore}, {Miller}, {Mills}, {Miraval}, {Moeyens}, {Moolekamp}, {Monet}, {Moniez}, {Monkewitz}, {Montgomery}, {Morrison}, {Mueller}, {Muller}, {Mu{\~n}oz Arancibia}, {Neill}, {Newbry}, {Nief}, {Nomerotski}, {Nordby}, {O'Connor}, {Oliver}, {Olivier}, {Olsen}, {O'Mullane}, {Ortiz}, {Osier}, {Owen}, {Pain}, {Palecek}, {Parejko}, {Parsons}, {Pease}, {Peterson}, {Peterson}, {Petravick}, {Libby Petrick}, {Petry}, {Pierfederici}, {Pietrowicz}, {Pike}, {Pinto}, {Plante}, {Plate}, {Plutchak}, {Price}, {Prouza}, {Radeka}, {Rajagopal}, {Rasmussen}, {Regnault}, {Reil}, {Reiss}, {Reuter}, {Ridgway}, {Riot}, {Ritz}, {Robinson}, {Roby}, {Roodman}, {Rosing}, {Roucelle}, {Rumore}, {Russo}, {Saha}, {Sassolas}, {Schalk}, {Schellart}, {Schindler}, {Schmidt}, {Schneider}, {Schneider}, {Schoening}, {Schumacher}, {Schwamb}, {Sebag}, {Selvy}, {Sembroski}, {Seppala}, {Serio}, {Serrano}, {Shaw}, {Shipsey}, {Sick}, {Silvestri},
  {Slater}, {Smith}, {Smith}, {Sobhani}, {Soldahl}, {Storrie-Lombardi}, {Stover}, {Strauss}, {Street}, {Stubbs}, {Sullivan}, {Sweeney}, {Swinbank}, {Szalay}, {Takacs}, {Tether}, {Thaler}, {Thayer}, {Thomas}, {Thornton}, {Thukral}, {Tice}, {Trilling}, {Turri}, {Van Berg}, {Vanden Berk}, {Vetter}, {Virieux}, {Vucina}, {Wahl}, {Walkowicz}, {Walsh}, {Walter}, {Wang}, {Wang}, {Warner}, {Wiecha}, {Willman}, {Winters}, {Wittman}, {Wolff}, {Wood-Vasey}, {Wu}, {Xin}, {Yoachim}, \& {Zhan}}]{2019ApJ...873..111I}
{Ivezi{\'c}}, {\v{Z}}., {Kahn}, S.~M., {Tyson}, J.~A., {et~al.} 2019, \href{http://dx.doi.org/10.3847/1538-4357/ab042c}{\color{blue}\apj}, \href{https://ui.adsabs.harvard.edu/abs/2019ApJ...873..111I}{873, 111}

\bibitem[{{Jamieson} {et~al.}(2024){Jamieson}, {Li}, {Villaescusa-Navarro}, {Ho}, \& {Spergel}}]{2024arXiv240807699J}
{Jamieson}, D., {Li}, Y., {Villaescusa-Navarro}, F., {Ho}, S., \& {Spergel}, D.~N. 2024, \href{https://ui.adsabs.harvard.edu/abs/2024arXiv240807699J}{\href{http://dx.doi.org/10.48550/arXiv.2408.07699}{\color{blue}arXiv e-prints}, arXiv:2408.07699}

\bibitem[{{Kodi Ramanah} {et~al.}(2020){Kodi Ramanah}, {Charnock}, {Villaescusa-Navarro}, \& {Wandelt}}]{2020MNRAS.495.4227K}
{Kodi Ramanah}, D., {Charnock}, T., {Villaescusa-Navarro}, F., \& {Wandelt}, B.~D. 2020, \href{http://dx.doi.org/10.1093/mnras/staa1428}{\color{blue}\mnras}, \href{https://ui.adsabs.harvard.edu/abs/2020MNRAS.495.4227K}{495, 4227}

\bibitem[{{Leclercq}(2018)}]{2018PhRvD..98f3511L}
{Leclercq}, F. 2018, \href{http://dx.doi.org/10.1103/PhysRevD.98.063511}{\color{blue}\prd}, \href{https://ui.adsabs.harvard.edu/abs/2018PhRvD..98f3511L}{98, 063511}

\bibitem[{{Lemos} {et~al.}(2021){Lemos}, {Raveri}, {Campos}, {Park}, {Chang}, {Weaverdyck}, {Huterer}, {Liddle}, {Blazek}, {Cawthon}, {Choi}, {DeRose}, {Dodelson}, {Doux}, {Gatti}, {Gruen}, {Harrison}, {Krause}, {Lahav}, {MacCrann}, {Muir}, {Prat}, {Rau}, {Rollins}, {Samuroff}, {Zuntz}, {Aguena}, {Allam}, {Annis}, {Avila}, {Bacon}, {Bernstein}, {Bertin}, {Brooks}, {Burke}, {Carnero Rosell}, {Carrasco Kind}, {Carretero}, {Castander}, {Conselice}, {Costanzi}, {Crocce}, {Pereira}, {Davis}, {De Vicente}, {Desai}, {Diehl}, {Doel}, {Eckert}, {Eifler}, {Elvin-Poole}, {Everett}, {Evrard}, {Ferrero}, {Fert{\'e}}, {Flaugher}, {Fosalba}, {Frieman}, {Garc{\'\i}a-Bellido}, {Gaztanaga}, {Gerdes}, {Giannantonio}, {Gruendl}, {Gschwend}, {Gutierrez}, {Hartley}, {Hinton}, {Hollowood}, {Honscheid}, {Hoyle}, {Huff}, {James}, {Jarvis}, {Lima}, {Maia}, {March}, {Marshall}, {Martini}, {Melchior}, {Menanteau}, {Miquel}, {Mohr}, {Morgan}, {Myles}, {Ogando}, {Palmese}, {Pandey}, {Paz-Chinch{\'o}n}, {Plazas Malag{\'o}n},
  {Rodriguez-Monroy}, {Roodman}, {Sanchez}, {Scarpine}, {Schubnell}, {Secco}, {Serrano}, {Sevilla-Noarbe}, {Smith}, {Soares-Santos}, {Suchyta}, {Swanson}, {Tarle}, {Thomas}, {To}, {Troxel}, {Varga}, {Weller}, {Wester}, \& {DES Collaboration}}]{2021MNRAS.505.6179L}
{Lemos}, P., {Raveri}, M., {Campos}, A., {et~al.} 2021, \href{http://dx.doi.org/10.1093/mnras/stab1670}{\color{blue}\mnras}, \href{https://ui.adsabs.harvard.edu/abs/2021MNRAS.505.6179L}{505, 6179}

\bibitem[{{Lewis}(2013)}]{2013PhRvD..87j3529L}
{Lewis}, A. 2013, \href{http://dx.doi.org/10.1103/PhysRevD.87.103529}{\color{blue}\prd}, \href{https://ui.adsabs.harvard.edu/abs/2013PhRvD..87j3529L}{87, 103529}

\bibitem[{{Lewis}(2019)}]{2019arXiv191013970L}
{Lewis}, A. 2019, \href{https://ui.adsabs.harvard.edu/abs/2019arXiv191013970L}{\href{http://dx.doi.org/10.48550/arXiv.1910.13970}{\color{blue}arXiv e-prints}, arXiv:1910.13970}

\bibitem[{{Lewis} {et~al.}(2000){Lewis}, {Challinor}, \& {Lasenby}}]{2000ApJ...538..473L}
{Lewis}, A., {Challinor}, A., \& {Lasenby}, A. 2000, \href{http://dx.doi.org/10.1086/309179}{\color{blue}\apj}, \href{https://ui.adsabs.harvard.edu/abs/2000ApJ...538..473L}{538, 473}

\bibitem[{{McEwen} {et~al.}(2021){McEwen}, {Wallis}, {Price}, \& {Spurio Mancini}}]{2021arXiv211112720M}
{McEwen}, J.~D., {Wallis}, C. G.~R., {Price}, M.~A., \& {Spurio Mancini}, A. 2021, \href{https://ui.adsabs.harvard.edu/abs/2021arXiv211112720M}{\href{http://dx.doi.org/10.48550/arXiv.2111.12720}{\color{blue}arXiv e-prints}, arXiv:2111.12720}

\bibitem[{{Mead} {et~al.}(2016){Mead}, {Heymans}, {Lombriser}, {Peacock}, {Steele}, \& {Winther}}]{2016MNRAS.459.1468M}
{Mead}, A.~J., {Heymans}, C., {Lombriser}, L., {et~al.} 2016, \href{http://dx.doi.org/10.1093/mnras/stw681}{\color{blue}\mnras}, \href{https://ui.adsabs.harvard.edu/abs/2016MNRAS.459.1468M}{459, 1468}

\bibitem[{{Mead} {et~al.}(2015){Mead}, {Peacock}, {Heymans}, {Joudaki}, \& {Heavens}}]{2015MNRAS.454.1958M}
{Mead}, A.~J., {Peacock}, J.~A., {Heymans}, C., {Joudaki}, S., \& {Heavens}, A.~F. 2015, \href{http://dx.doi.org/10.1093/mnras/stv2036}{\color{blue}\mnras}, \href{https://ui.adsabs.harvard.edu/abs/2015MNRAS.454.1958M}{454, 1958}

\bibitem[{{Mootoovaloo} {et~al.}(2020){Mootoovaloo}, {Heavens}, {Jaffe}, \& {Leclercq}}]{2020MNRAS.497.2213M}
{Mootoovaloo}, A., {Heavens}, A.~F., {Jaffe}, A.~H., \& {Leclercq}, F. 2020, \href{http://dx.doi.org/10.1093/mnras/staa2102}{\color{blue}\mnras}, \href{https://ui.adsabs.harvard.edu/abs/2020MNRAS.497.2213M}{497, 2213}

\bibitem[{{Mootoovaloo} {et~al.}(2022){Mootoovaloo}, {Jaffe}, {Heavens}, \& {Leclercq}}]{2022A&C....3800508M}
{Mootoovaloo}, A., {Jaffe}, A.~H., {Heavens}, A.~F., \& {Leclercq}, F. 2022, \href{http://dx.doi.org/10.1016/j.ascom.2021.100508}{\color{blue}Astronomy and Computing}, \href{https://ui.adsabs.harvard.edu/abs/2022A&C....3800508M}{38, 100508}

\bibitem[{{Mootoovaloo} {et~al.}(2024){Mootoovaloo}, {Ruiz-Zapatero}, {Garc{\'\i}a-Garc{\'\i}a}, \& {Alonso}}]{2024arXiv240604725M}
{Mootoovaloo}, A., {Ruiz-Zapatero}, J., {Garc{\'\i}a-Garc{\'\i}a}, C., \& {Alonso}, D. 2024, \href{https://ui.adsabs.harvard.edu/abs/2024arXiv240604725M}{\href{http://dx.doi.org/10.48550/arXiv.2406.04725}{\color{blue}arXiv e-prints}, arXiv:2406.04725}

\bibitem[{{Paszke} {et~al.}(2019){Paszke}, {Gross}, {Massa}, {Lerer}, {Bradbury}, {Chanan}, {Killeen}, {Lin}, {Gimelshein}, {Antiga}, {Desmaison}, {K{\"o}pf}, {Yang}, {DeVito}, {Raison}, {Tejani}, {Chilamkurthy}, {Steiner}, {Fang}, {Bai}, \& {Chintala}}]{2019arXiv191201703P}
{Paszke}, A., {Gross}, S., {Massa}, F., {et~al.} 2019, \href{https://ui.adsabs.harvard.edu/abs/2019arXiv191201703P}{\href{http://dx.doi.org/10.48550/arXiv.1912.01703}{\color{blue}arXiv e-prints}, arXiv:1912.01703}

\bibitem[{{Planck Collaboration} {et~al.}(2020{\natexlab{a}}){Planck Collaboration}, {Aghanim}, {Akrami}, {Ashdown}, {Aumont}, {Baccigalupi}, {Ballardini}, {Banday}, {Barreiro}, {Bartolo}, {Basak}, {Battye}, {Benabed}, {Bernard}, {Bersanelli}, {Bielewicz}, {Bock}, {Bond}, {Borrill}, {Bouchet}, {Boulanger}, {Bucher}, {Burigana}, {Butler}, {Calabrese}, {Cardoso}, {Carron}, {Challinor}, {Chiang}, {Chluba}, {Colombo}, {Combet}, {Contreras}, {Crill}, {Cuttaia}, {de Bernardis}, {de Zotti}, {Delabrouille}, {Delouis}, {Di Valentino}, {Diego}, {Dor{\'e}}, {Douspis}, {Ducout}, {Dupac}, {Dusini}, {Efstathiou}, {Elsner}, {En{\ss}lin}, {Eriksen}, {Fantaye}, {Farhang}, {Fergusson}, {Fernandez-Cobos}, {Finelli}, {Forastieri}, {Frailis}, {Fraisse}, {Franceschi}, {Frolov}, {Galeotta}, {Galli}, {Ganga}, {G{\'e}nova-Santos}, {Gerbino}, {Ghosh}, {Gonz{\'a}lez-Nuevo}, {G{\'o}rski}, {Gratton}, {Gruppuso}, {Gudmundsson}, {Hamann}, {Handley}, {Hansen}, {Herranz}, {Hildebrandt}, {Hivon}, {Huang}, {Jaffe}, {Jones}, {Karakci},
  {Keih{\"a}nen}, {Keskitalo}, {Kiiveri}, {Kim}, {Kisner}, {Knox}, {Krachmalnicoff}, {Kunz}, {Kurki-Suonio}, {Lagache}, {Lamarre}, {Lasenby}, {Lattanzi}, {Lawrence}, {Le Jeune}, {Lemos}, {Lesgourgues}, {Levrier}, {Lewis}, {Liguori}, {Lilje}, {Lilley}, {Lindholm}, {L{\'o}pez-Caniego}, {Lubin}, {Ma}, {Mac{\'\i}as-P{\'e}rez}, {Maggio}, {Maino}, {Mandolesi}, {Mangilli}, {Marcos-Caballero}, {Maris}, {Martin}, {Martinelli}, {Mart{\'\i}nez-Gonz{\'a}lez}, {Matarrese}, {Mauri}, {McEwen}, {Meinhold}, {Melchiorri}, {Mennella}, {Migliaccio}, {Millea}, {Mitra}, {Miville-Desch{\^e}nes}, {Molinari}, {Montier}, {Morgante}, {Moss}, {Natoli}, {N{\o}rgaard-Nielsen}, {Pagano}, {Paoletti}, {Partridge}, {Patanchon}, {Peiris}, {Perrotta}, {Pettorino}, {Piacentini}, {Polastri}, {Polenta}, {Puget}, {Rachen}, {Reinecke}, {Remazeilles}, {Renzi}, {Rocha}, {Rosset}, {Roudier}, {Rubi{\~n}o-Mart{\'\i}n}, {Ruiz-Granados}, {Salvati}, {Sandri}, {Savelainen}, {Scott}, {Shellard}, {Sirignano}, {Sirri}, {Spencer}, {Sunyaev}, {Suur-Uski},
  {Tauber}, {Tavagnacco}, {Tenti}, {Toffolatti}, {Tomasi}, {Trombetti}, {Valenziano}, {Valiviita}, {Van Tent}, {Vibert}, {Vielva}, {Villa}, {Vittorio}, {Wandelt}, {Wehus}, {White}, {White}, {Zacchei}, \& {Zonca}}]{2020A&A...641A...6P}
{Planck Collaboration}, {Aghanim}, N., {Akrami}, Y., {et~al.} 2020{\natexlab{a}}, \href{http://dx.doi.org/10.1051/0004-6361/201833910}{\color{blue}\aap}, \href{https://ui.adsabs.harvard.edu/abs/2020A&A...641A...6P}{641, A6}

\bibitem[{{Planck Collaboration} {et~al.}(2020{\natexlab{b}}){Planck Collaboration}, {Aghanim}, {Akrami}, {Ashdown}, {Aumont}, {Baccigalupi}, {Ballardini}, {Banday}, {Barreiro}, {Bartolo}, {Basak}, {Battye}, {Benabed}, {Bernard}, {Bersanelli}, {Bielewicz}, {Bock}, {Bond}, {Borrill}, {Bouchet}, {Boulanger}, {Bucher}, {Burigana}, {Butler}, {Calabrese}, {Cardoso}, {Carron}, {Challinor}, {Chiang}, {Chluba}, {Colombo}, {Combet}, {Contreras}, {Crill}, {Cuttaia}, {de Bernardis}, {de Zotti}, {Delabrouille}, {Delouis}, {Di Valentino}, {Diego}, {Dor{\'e}}, {Douspis}, {Ducout}, {Dupac}, {Dusini}, {Efstathiou}, {Elsner}, {En{\ss}lin}, {Eriksen}, {Fantaye}, {Farhang}, {Fergusson}, {Fernandez-Cobos}, {Finelli}, {Forastieri}, {Frailis}, {Fraisse}, {Franceschi}, {Frolov}, {Galeotta}, {Galli}, {Ganga}, {G{\'e}nova-Santos}, {Gerbino}, {Ghosh}, {Gonz{\'a}lez-Nuevo}, {G{\'o}rski}, {Gratton}, {Gruppuso}, {Gudmundsson}, {Hamann}, {Handley}, {Hansen}, {Herranz}, {Hildebrandt}, {Hivon}, {Huang}, {Jaffe}, {Jones}, {Karakci},
  {Keih{\"a}nen}, {Keskitalo}, {Kiiveri}, {Kim}, {Kisner}, {Knox}, {Krachmalnicoff}, {Kunz}, {Kurki-Suonio}, {Lagache}, {Lamarre}, {Lasenby}, {Lattanzi}, {Lawrence}, {Le Jeune}, {Lemos}, {Lesgourgues}, {Levrier}, {Lewis}, {Liguori}, {Lilje}, {Lilley}, {Lindholm}, {L{\'o}pez-Caniego}, {Lubin}, {Ma}, {Mac{\'\i}as-P{\'e}rez}, {Maggio}, {Maino}, {Mandolesi}, {Mangilli}, {Marcos-Caballero}, {Maris}, {Martin}, {Martinelli}, {Mart{\'\i}nez-Gonz{\'a}lez}, {Matarrese}, {Mauri}, {McEwen}, {Meinhold}, {Melchiorri}, {Mennella}, {Migliaccio}, {Millea}, {Mitra}, {Miville-Desch{\^e}nes}, {Molinari}, {Montier}, {Morgante}, {Moss}, {Natoli}, {N{\o}rgaard-Nielsen}, {Pagano}, {Paoletti}, {Partridge}, {Patanchon}, {Peiris}, {Perrotta}, {Pettorino}, {Piacentini}, {Polastri}, {Polenta}, {Puget}, {Rachen}, {Reinecke}, {Remazeilles}, {Renzi}, {Rocha}, {Rosset}, {Roudier}, {Rubi{\~n}o-Mart{\'\i}n}, {Ruiz-Granados}, {Salvati}, {Sandri}, {Savelainen}, {Scott}, {Shellard}, {Sirignano}, {Sirri}, {Spencer}, {Sunyaev}, {Suur-Uski},
  {Tauber}, {Tavagnacco}, {Tenti}, {Toffolatti}, {Tomasi}, {Trombetti}, {Valenziano}, {Valiviita}, {Van Tent}, {Vibert}, {Vielva}, {Villa}, {Vittorio}, {Wandelt}, {Wehus}, {White}, {White}, {Zacchei}, \& {Zonca}}]{Planck:2018lbu}
{Planck Collaboration}, {Aghanim}, N., {Akrami}, Y., {et~al.} 2020{\natexlab{b}}, \href{http://dx.doi.org/10.1051/0004-6361/201833910}{\color{blue}\aap}, \href{https://ui.adsabs.harvard.edu/abs/2020A&A...641A...6P}{641, A6}

\bibitem[{{Planck Collaboration} {et~al.}(2020{\natexlab{c}}){Planck Collaboration}, {Aghanim}, {Akrami}, {Ashdown}, {Aumont}, {Baccigalupi}, {Ballardini}, {Banday}, {Barreiro}, {Bartolo}, {Basak}, {Benabed}, {Bernard}, {Bersanelli}, {Bielewicz}, {Bock}, {Bond}, {Borrill}, {Bouchet}, {Boulanger}, {Bucher}, {Burigana}, {Butler}, {Calabrese}, {Cardoso}, {Carron}, {Casaponsa}, {Challinor}, {Chiang}, {Colombo}, {Combet}, {Crill}, {Cuttaia}, {de Bernardis}, {de Rosa}, {de Zotti}, {Delabrouille}, {Delouis}, {Di Valentino}, {Diego}, {Dor{\'e}}, {Douspis}, {Ducout}, {Dupac}, {Dusini}, {Efstathiou}, {Elsner}, {En{\ss}lin}, {Eriksen}, {Fantaye}, {Fernandez-Cobos}, {Finelli}, {Frailis}, {Fraisse}, {Franceschi}, {Frolov}, {Galeotta}, {Galli}, {Ganga}, {G{\'e}nova-Santos}, {Gerbino}, {Ghosh}, {Giraud-H{\'e}raud}, {Gonz{\'a}lez-Nuevo}, {G{\'o}rski}, {Gratton}, {Gruppuso}, {Gudmundsson}, {Hamann}, {Handley}, {Hansen}, {Herranz}, {Hivon}, {Huang}, {Jaffe}, {Jones}, {Keih{\"a}nen}, {Keskitalo}, {Kiiveri}, {Kim}, {Kisner},
  {Krachmalnicoff}, {Kunz}, {Kurki-Suonio}, {Lagache}, {Lamarre}, {Lasenby}, {Lattanzi}, {Lawrence}, {Le Jeune}, {Levrier}, {Lewis}, {Liguori}, {Lilje}, {Lilley}, {Lindholm}, {L{\'o}pez-Caniego}, {Lubin}, {Ma}, {Mac{\'\i}as-P{\'e}rez}, {Maggio}, {Maino}, {Mandolesi}, {Mangilli}, {Marcos-Caballero}, {Maris}, {Martin}, {Mart{\'\i}nez-Gonz{\'a}lez}, {Matarrese}, {Mauri}, {McEwen}, {Meinhold}, {Melchiorri}, {Mennella}, {Migliaccio}, {Millea}, {Miville-Desch{\^e}nes}, {Molinari}, {Moneti}, {Montier}, {Morgante}, {Moss}, {Natoli}, {N{\o}rgaard-Nielsen}, {Pagano}, {Paoletti}, {Partridge}, {Patanchon}, {Peiris}, {Perrotta}, {Pettorino}, {Piacentini}, {Polenta}, {Puget}, {Rachen}, {Reinecke}, {Remazeilles}, {Renzi}, {Rocha}, {Rosset}, {Roudier}, {Rubi{\~n}o-Mart{\'\i}n}, {Ruiz-Granados}, {Salvati}, {Sandri}, {Savelainen}, {Scott}, {Shellard}, {Sirignano}, {Sirri}, {Spencer}, {Sunyaev}, {Suur-Uski}, {Tauber}, {Tavagnacco}, {Tenti}, {Toffolatti}, {Tomasi}, {Trombetti}, {Valiviita}, {Van Tent}, {Vielva}, {Villa},
  {Vittorio}, {Wandelt}, {Wehus}, {Zacchei}, \& {Zonca}}]{Planck:2019nip}
{Planck Collaboration}, {Aghanim}, N., {Akrami}, Y., {et~al.} 2020{\natexlab{c}}, \href{http://dx.doi.org/10.1051/0004-6361/201936386}{\color{blue}\aap}, \href{https://ui.adsabs.harvard.edu/abs/2020A&A...641A...5P}{641, A5}

\bibitem[{{Planck Collaboration} {et~al.}(2020{\natexlab{d}}){Planck Collaboration}, {Aghanim}, {Akrami}, {Ashdown}, {Aumont}, {Baccigalupi}, {Ballardini}, {Banday}, {Barreiro}, {Bartolo}, {Basak}, {Benabed}, {Bernard}, {Bersanelli}, {Bielewicz}, {Bock}, {Bond}, {Borrill}, {Bouchet}, {Boulanger}, {Bucher}, {Burigana}, {Calabrese}, {Cardoso}, {Carron}, {Challinor}, {Chiang}, {Colombo}, {Combet}, {Crill}, {Cuttaia}, {de Bernardis}, {de Zotti}, {Delabrouille}, {Di Valentino}, {Diego}, {Dor{\'e}}, {Douspis}, {Ducout}, {Dupac}, {Efstathiou}, {Elsner}, {En{\ss}lin}, {Eriksen}, {Fantaye}, {Fernandez-Cobos}, {Finelli}, {Forastieri}, {Frailis}, {Fraisse}, {Franceschi}, {Frolov}, {Galeotta}, {Galli}, {Ganga}, {G{\'e}nova-Santos}, {Gerbino}, {Ghosh}, {Gonz{\'a}lez-Nuevo}, {G{\'o}rski}, {Gratton}, {Gruppuso}, {Gudmundsson}, {Hamann}, {Handley}, {Hansen}, {Herranz}, {Hivon}, {Huang}, {Jaffe}, {Jones}, {Karakci}, {Keih{\"a}nen}, {Keskitalo}, {Kiiveri}, {Kim}, {Knox}, {Krachmalnicoff}, {Kunz}, {Kurki-Suonio}, {Lagache},
  {Lamarre}, {Lasenby}, {Lattanzi}, {Lawrence}, {Le Jeune}, {Levrier}, {Lewis}, {Liguori}, {Lilje}, {Lindholm}, {L{\'o}pez-Caniego}, {Lubin}, {Ma}, {Mac{\'\i}as-P{\'e}rez}, {Maggio}, {Maino}, {Mandolesi}, {Mangilli}, {Marcos-Caballero}, {Maris}, {Martin}, {Mart{\'\i}nez-Gonz{\'a}lez}, {Matarrese}, {Mauri}, {McEwen}, {Melchiorri}, {Mennella}, {Migliaccio}, {Miville-Desch{\^e}nes}, {Molinari}, {Moneti}, {Montier}, {Morgante}, {Moss}, {Natoli}, {Pagano}, {Paoletti}, {Partridge}, {Patanchon}, {Perrotta}, {Pettorino}, {Piacentini}, {Polastri}, {Polenta}, {Puget}, {Rachen}, {Reinecke}, {Remazeilles}, {Renzi}, {Rocha}, {Rosset}, {Roudier}, {Rubi{\~n}o-Mart{\'\i}n}, {Ruiz-Granados}, {Salvati}, {Sandri}, {Savelainen}, {Scott}, {Sirignano}, {Sunyaev}, {Suur-Uski}, {Tauber}, {Tavagnacco}, {Tenti}, {Toffolatti}, {Tomasi}, {Trombetti}, {Valiviita}, {Van Tent}, {Vielva}, {Villa}, {Vittorio}, {Wandelt}, {Wehus}, {White}, {White}, {Zacchei}, \& {Zonca}}]{1807.06210}
{Planck Collaboration}, {Aghanim}, N., {Akrami}, Y., {et~al.} 2020{\natexlab{d}}, \href{http://dx.doi.org/10.1051/0004-6361/201833886}{\color{blue}\aap}, \href{https://ui.adsabs.harvard.edu/abs/2020A&A...641A...8P}{641, A8}

\bibitem[{{Polanska} {et~al.}(2024){Polanska}, {Price}, {Piras}, {Spurio Mancini}, \& {McEwen}}]{2024arXiv240505969P}
{Polanska}, A., {Price}, M.~A., {Piras}, D., {Spurio Mancini}, A., \& {McEwen}, J.~D. 2024, \href{https://ui.adsabs.harvard.edu/abs/2024arXiv240505969P}{\href{http://dx.doi.org/10.48550/arXiv.2405.05969}{\color{blue}arXiv e-prints}, arXiv:2405.05969}

\bibitem[{{Reback} {et~al.}(2020){Reback}, {McKinney}, {Jbrockmendel}, {Van Den Bossche}, {Augspurger}, {Cloud}, {Gfyoung}, {Sinhrks}, {Klein}, {Roeschke}, {Tratner}, {She}, {Ayd}, {Hawkins}, {Petersen}, {Schendel}, {Hayden}, {Garcia}, {Jancauskas}, {MomIsBestFriend}, {Battiston}, {Seabold}, {Chris-B1}, {H-Vetinari}, {Hoyer}, {Overmeire}, {Alimcmaster1}, {Mehyar}, {Whelan}, \& {Kluyver}}]{jeff_reback_2020_3630805}
{Reback}, J., {McKinney}, W., {Jbrockmendel}, {et~al.} 2020, {pandas-dev/pandas: Pandas 1.0.0}

\bibitem[{Rizzo \& Sz{\'e}kely(2016)}]{rizzo2016energy}
Rizzo, M.~L. \& Sz{\'e}kely, G.~J. 2016, wiley interdisciplinary reviews: Computational statistics, 8, 8

\bibitem[{{Rozo} {et~al.}(2016){Rozo}, {Rykoff}, {Abate}, {Bonnett}, {Crocce}, {Davis}, {Hoyle}, {Leistedt}, {Peiris}, {Wechsler}, {Abbott}, {Abdalla}, {Banerji}, {Bauer}, {Benoit-L{\'e}vy}, {Bernstein}, {Bertin}, {Brooks}, {Buckley-Geer}, {Burke}, {Capozzi}, {Rosell}, {Carollo}, {Kind}, {Carretero}, {Castander}, {Childress}, {Cunha}, {D'Andrea}, {Davis}, {DePoy}, {Desai}, {Diehl}, {Dietrich}, {Doel}, {Eifler}, {Evrard}, {Neto}, {Flaugher}, {Fosalba}, {Frieman}, {Gaztanaga}, {Gerdes}, {Glazebrook}, {Gruen}, {Gruendl}, {Honscheid}, {James}, {Jarvis}, {Kim}, {Kuehn}, {Kuropatkin}, {Lahav}, {Lidman}, {Lima}, {Maia}, {March}, {Martini}, {Melchior}, {Miller}, {Miquel}, {Mohr}, {Nichol}, {Nord}, {O'Neill}, {Ogando}, {Plazas}, {Romer}, {Roodman}, {Sako}, {Sanchez}, {Santiago}, {Schubnell}, {Sevilla-Noarbe}, {Smith}, {Soares-Santos}, {Sobreira}, {Suchyta}, {Swanson}, {Thaler}, {Thomas}, {Uddin}, {Vikram}, {Walker}, {Wester}, {Zhang}, \& {da Costa}}]{1507.05460}
{Rozo}, E., {Rykoff}, E.~S., {Abate}, A., {et~al.} 2016, \href{http://dx.doi.org/10.1093/mnras/stw1281}{\color{blue}\mnras}, \href{https://ui.adsabs.harvard.edu/abs/2016MNRAS.461.1431R}{461, 1431}

\bibitem[{{Ruiz-Zapatero} {et~al.}(2023){Ruiz-Zapatero}, {Hadzhiyska}, {Alonso}, {Ferreira}, {Garc{\'\i}a-Garc{\'\i}a}, \& {Mootoovaloo}}]{2301.11978}
{Ruiz-Zapatero}, J., {Hadzhiyska}, B., {Alonso}, D., {et~al.} 2023, \href{http://dx.doi.org/10.1093/mnras/stad1192}{\color{blue}\mnras}, \href{https://ui.adsabs.harvard.edu/abs/2023MNRAS.522.5037R}{522, 5037}

\bibitem[{{Spurio Mancini} {et~al.}(2022){Spurio Mancini}, {Piras}, {Alsing}, {Joachimi}, \& {Hobson}}]{2022MNRAS.511.1771S}
{Spurio Mancini}, A., {Piras}, D., {Alsing}, J., {Joachimi}, B., \& {Hobson}, M.~P. 2022, \href{http://dx.doi.org/10.1093/mnras/stac064}{\color{blue}\mnras}, \href{https://ui.adsabs.harvard.edu/abs/2022MNRAS.511.1771S}{511, 1771}

\bibitem[{{Srinivasan} {et~al.}(2024){Srinivasan}, {Crisostomi}, {Trotta}, {Barausse}, \& {Breschi}}]{2024arXiv240412294S}
{Srinivasan}, R., {Crisostomi}, M., {Trotta}, R., {Barausse}, E., \& {Breschi}, M. 2024, \href{https://ui.adsabs.harvard.edu/abs/2024arXiv240412294S}{\href{http://dx.doi.org/10.48550/arXiv.2404.12294}{\color{blue}arXiv e-prints}, arXiv:2404.12294}

\bibitem[{{Takahashi} {et~al.}(2012){Takahashi}, {Sato}, {Nishimichi}, {Taruya}, \& {Oguri}}]{1208.2701}
{Takahashi}, R., {Sato}, M., {Nishimichi}, T., {Taruya}, A., \& {Oguri}, M. 2012, \href{http://dx.doi.org/10.1088/0004-637X/761/2/152}{\color{blue}\apj}, \href{https://ui.adsabs.harvard.edu/abs/2012ApJ...761..152T}{761, 152}

\bibitem[{{Torrado} \& {Lewis}(2021)}]{2005.05290}
{Torrado}, J. \& {Lewis}, A. 2021, \href{http://dx.doi.org/10.1088/1475-7516/2021/05/057}{\color{blue}\jcap}, \href{https://ui.adsabs.harvard.edu/abs/2021JCAP...05..057T}{2021, 057}

\bibitem[{Tresp(2000)}]{tresp2000bayesian}
Tresp, V. 2000, Neural computation, 12, 12

\bibitem[{{Tr{\"o}ster} {et~al.}(2019){Tr{\"o}ster}, {Ferguson}, {Harnois-D{\'e}raps}, \& {McCarthy}}]{2019MNRAS.487L..24T}
{Tr{\"o}ster}, T., {Ferguson}, C., {Harnois-D{\'e}raps}, J., \& {McCarthy}, I.~G. 2019, \href{http://dx.doi.org/10.1093/mnrasl/slz075}{\color{blue}\mnras}, \href{https://ui.adsabs.harvard.edu/abs/2019MNRAS.487L..24T}{487, L24}

\bibitem[{{Virtanen} {et~al.}(2020){Virtanen}, {Gommers}, {Oliphant}, {Haberland}, {Reddy}, {Cournapeau}, {Burovski}, {Peterson}, {Weckesser}, {Bright}, {van der Walt}, {Brett}, {Wilson}, {Millman}, {Mayorov}, {Nelson}, {Jones}, {Kern}, {Larson}, {Carey}, {Polat}, {Feng}, {Moore}, {VanderPlas}, {Laxalde}, {Perktold}, {Cimrman}, {Henriksen}, {Quintero}, {Harris}, {Archibald}, {Ribeiro}, {Pedregosa}, {van Mulbregt}, \& {SciPy 1. 0 Contributors}}]{2020NatMe..17..261V}
{Virtanen}, P., {Gommers}, R., {Oliphant}, T.~E., {et~al.} 2020, \href{http://dx.doi.org/10.1038/s41592-019-0686-2}{\color{blue}Nature Methods}, \href{https://ui.adsabs.harvard.edu/abs/2020NatMe..17..261V}{17, 261}

\bibitem[{Webb(2022)}]{WebbFlowtorch}
Webb, S. 2022, FlowTorch, Github

\bibitem[{{Wu} {et~al.}(2024){Wu}, {Imbiriba}, {Elvira}, \& {Closas}}]{2024ITSP...72..275W}
{Wu}, P., {Imbiriba}, T., {Elvira}, V., \& {Closas}, P. 2024, \href{http://dx.doi.org/10.1109/TSP.2023.3343564}{\color{blue}IEEE Transactions on Signal Processing}, \href{https://ui.adsabs.harvard.edu/abs/2024ITSP...72..275W}{72, 275}

\bibitem[{Yadan(2019)}]{Yadan2019Hydra}
Yadan, O. 2019, Hydra - A framework for elegantly configuring complex applications, Github

\bibitem[{{Zuntz} {et~al.}(2018){Zuntz}, {Sheldon}, {Samuroff}, {Troxel}, {Jarvis}, {MacCrann}, {Gruen}, {Prat}, {S{\'a}nchez}, {Choi}, {Bridle}, {Bernstein}, {Dodelson}, {Drlica-Wagner}, {Fang}, {Gruendl}, {Hoyle}, {Huff}, {Jain}, {Kirk}, {Kacprzak}, {Krawiec}, {Plazas}, {Rollins}, {Rykoff}, {Sevilla-Noarbe}, {Soergel}, {Varga}, {Abbott}, {Abdalla}, {Allam}, {Annis}, {Bechtol}, {Benoit-L{\'e}vy}, {Bertin}, {Buckley-Geer}, {Burke}, {Carnero Rosell}, {Carrasco Kind}, {Carretero}, {Castander}, {Crocce}, {Cunha}, {D'Andrea}, {da Costa}, {Davis}, {Desai}, {Diehl}, {Dietrich}, {Doel}, {Eifler}, {Estrada}, {Evrard}, {Fausti Neto}, {Fernandez}, {Flaugher}, {Fosalba}, {Frieman}, {Garc{\'\i}a-Bellido}, {Gaztanaga}, {Gerdes}, {Giannantonio}, {Gschwend}, {Gutierrez}, {Hartley}, {Honscheid}, {James}, {Jeltema}, {Johnson}, {Johnson}, {Kuehn}, {Kuhlmann}, {Kuropatkin}, {Lahav}, {Li}, {Lima}, {Maia}, {March}, {Martini}, {Melchior}, {Menanteau}, {Miller}, {Miquel}, {Mohr}, {Neilsen}, {Nichol}, {Ogando}, {Roe}, {Romer},
  {Roodman}, {Sanchez}, {Scarpine}, {Schindler}, {Schubnell}, {Smith}, {Smith}, {Soares-Santos}, {Sobreira}, {Suchyta}, {Swanson}, {Tarle}, {Thomas}, {Tucker}, {Vikram}, {Walker}, {Wechsler}, {Zhang}, \& {DES Collaboration}}]{1708.01533}
{Zuntz}, J., {Sheldon}, E., {Samuroff}, S., {et~al.} 2018, \href{http://dx.doi.org/10.1093/mnras/sty2219}{\color{blue}\mnras}, \href{https://ui.adsabs.harvard.edu/abs/2018MNRAS.481.1149Z}{481, 1149}

\end{thebibliography}
